\documentclass[11pt,fleqn]{article}
\pdfoutput=1
\usepackage{amsmath}
\usepackage{amsthm}
\usepackage{amsfonts}
\usepackage{amssymb}
\usepackage{mathtools}
\usepackage{fullpage}
\usepackage{color}
\usepackage{times}
\usepackage{graphicx}
\usepackage{float}

\usepackage{natbib}
\begin{document}
\renewcommand{\thefootnote}{\fnsymbol{footnote}}
\title{Forward problem for Love and quasi-Rayleigh waves:\\
Exact dispersion relations and their sensitivities}
\author{
David R. Dalton\footnote{
Department of Earth Sciences, Memorial University of Newfoundland,
{\tt dalton.nfld@gmail.com}}\,,
Michael A. Slawinski\footnote{
Department of Earth Sciences, Memorial University of Newfoundland,
{\tt mslawins@mac.com}}\,,
Piotr Stachura\footnote{
Faculty of Applied Informatics and Mathematics, SGGW, 
{\tt  piotr\_stachura1@sggw.pl}}\,,
Theodore Stanoev\footnote{
Department of Earth Sciences, Memorial University of Newfoundland,
{\tt theodore.stanoev@gmail.com}}
}
\date{\today}
\maketitle
\renewcommand{\thefootnote}{\arabic{footnote}}
\setcounter{footnote}{0}
\section*{Abstract}
We examine two types of guided waves: the Love and the quasi-Rayleigh waves.
Both waves propagate in the same model of an elastic isotropic layer above an elastic isotropic halfspace.  
From their dispersion relations, we calculate their speeds as functions of the elasticity parameters, mass densities, frequency and layer thickness.
We examine the sensitivity of these relations to the model and wave properties.
%%%%%%%%%%%%%%%%%%%%%%%%%%%%
\section{Introduction}
%%%%%%%%%%%%%%%%%%%%%%%%%%%%
In this paper, we examine the forward problem for two types of guided waves that propagate within the same model.
This model consists of an elastic isotropic layer over an elastic isotropic halfspace.
Given properties of the model---namely, its elasticity parameters, mass densities of both media, as well as the thickness of the layer---and the frequency of the signal, we derive the speeds of the waves that correspond to different modes for either wave.

Each wave has an infinite number of modes, and each mode propagates with a different speed.
For a given frequency, there is a finite number of modes, and hence, speeds.

On the surface, the two waves exhibit displacements that are perpendicular to each other.
The displacements of one wave, which is the Love wave, are in the horizontal plane and perpendicular to the direction of propagation.
The displacements of the other wave, which is the quasi-Rayleigh wave, are in the vertical plane, and---on the surface---are parallel to the direction of propagation.

We refer to the latter wave as quasi-Rayleigh, since it shares many similarities with the standard Rayleigh wave, but is not restricted to the halfspace.   
In literature, the quasi-Rayleigh wave has been also referred to as Rayleigh-type wave, Rayleigh-like wave, generalized Rayleigh wave, Rayleigh-Lamb wave and Rayleigh wave in inhomogeneous media.

Seismological information, such as wave speeds measured on the surface, might allow us to infer properties of the subsurface.
To gain an insight into such an inverse problem we examine details of the forward one.
Wave speeds corresponding to different modes of either wave provide unique information about certain model properties and redundant information of other properties; the latter quality might allow one to examine the reliability of a solution of the inverse problem.

In formulating and deriving the forward-problem expressions---and motivated by both accuracy of modern seismic measurements and availability of computational tools---we proceed without any simplifying assumptions beyond the standard approach of elasticity theory in isotropic media.
As far as we could ascertain, this is not the case of other studies presented in literature.
We examine the sensitivity of the fundamental expressions, which are the dispersion relations of the Love and quasi-Rayleigh waves, to such quantities as elasticity parameters and frequencies.

%%%%%%%%%%%%%%%%%%%%%%%%%%%%
\section{Literature review}
%%%%%%%%%%%%%%%%%%%%%%%%%%%%
Already \citet{Love1911} and~\citet{Rudzki1912} presented theoretical examinations of surface waves. 
\citet{Stoneley1934} analyzed transmission of Rayleigh waves in a heterogeneous medium with a constant density and with a rigidity varying linearly with depth.
\citet{Sezawa1927}, \citet{Lee1932} and~\citet{Fu1946} also analyzed the problem of an elastic plate above an elastic halfspace.
\citet{Newlands1950} discussed further Rayleigh waves in a two-layer heterogeneous medium.
\citet{Haskell1953} applied the matrix formulation of \citet{Thomson1950} to the dispersion of surface waves on multilayered media.
\citet{Anderson1961} analyzed dispersive properties of transversely isotropic media.  
\citet{Tiersten1969} examined elastic surface waves guided by thin films.  
\citet{Crampin1970} extended the Thomson-Haskell matrix formulation to anisotropic layers.
\mbox{\citet{Osipov1970}} studied Rayleigh-type waves in an anisotropic halfspace.     
\citet{Coste1997} approximated dispersion formu{l\ae} for Rayleigh-like waves in a layered medium.
\citet{KeEtAl2011} discussed two modified Thomson-Haskell matrix methods and associated accelerated root-searching schemes.
\citet{LiuFan2012} described the characteristics of high-frequency Rayleigh waves in a stratified halfspace.
\citet{VinhLinh2013} analyzed generalized Rayleigh waves in prestressed compressible elastic solids.
\citet{LaiEtAl2014} derived a relation for the effective phase velocity of Rayleigh waves in a vertically inhomogeneous isotropic elastic halfspace.

Quasi-Rayleigh waves within the model discussed herein are reviewed by~\citet[Section~10.4]{Udias1999}.
Unlike \citet{Udias1999}, we do not restrict Poisson's ratio in the layer and in the halfspace to be $1/4$\,.  
\cite{Fu1946} makes certain simplifying assumptions prior to calculations, which we do not. 
Such an approach allows us to examine details of the forward problem and sets the stage for a further investigation.
%%%%%%%%%%%%%%%%%%%%%%%%%%%%
\section{Love waves}
\label{sec:LovWav}
%%%%%%%%%%%%%%%%%%%%%%%%%%%%
\subsection{Material properties}
%%%%%%%%%%%%%%%%%%%%%%%%%%%%
We wish to examine guided $SH$ waves in an elastic layer over an elastic halfspace.
For this purpose, we consider an elastic layer with mass density,~$\rho^u$, elasticity parameter,~$C_{44}^u$\,, and hence $S$-wave propagation speed,~$\beta^u=\sqrt{C_{44}^u/\rho^u\strut}$\,.
Also, we consider an elastic halfspace with $\rho^d$, $C_{44}^d$ and $\beta^d=\sqrt{C_{44}^d/\rho^d}$.
We set the coordinate system in such a manner that the surface is at $x_3=0$ and the interface is at $x_3=Z$, with the~$x_3$-axis positive downwards.

Herein, we consider the $SH$ waves propagating only in the $x_1$-direction. 
Hence, two components of the displacement vector, $u_1=u_3=0$\,, for both the layer and the halfspace. We write the nonzero components of~$\bf u$ as
\begin{equation}
\label{eq:LoveAnsatzU}
u_2^u(x_1,x_3,t) =
A^u(x_3)\exp[\iota(\kappa x_1-\omega t)]\,,
\end{equation}
%and
\begin{equation}
\label{eq:LoveAnsatzD}
u_2^d(x_1,x_3,t) = 
A^d(x_3)\exp[\iota(\kappa x_1-\omega t)]\,,
\end{equation}
which are plane $SH$ waves propagating in the $x_1$-direction in the layer and halfspace, respectively.
The amplitudes of these waves are functions of $x_3$.

Since the $SH$ wave is decoupled from the $P$ and $SV$ waves, it suffices to consider its displacements.
We do not need to consider its scalar and vector potentials to examine interactions with other waves, which we do for the quasi-Rayleigh wave in Section~\ref{sec:qRwaves}.
%%%%%%%%%%%%%%%%%%%%%%%%%%%%
\subsection{{\boldmath $SH$} wave equation and its solutions}
\label{sub:DerSH}
%%%%%%%%%%%%%%%%%%%%%%%%%%%%
The wave equation that corresponds to the $SH$ waves propagating in the $x_1$-direction is
\begin{equation*}
\frac{\partial^2u_2^\mu}{\partial x_1^2} +
\frac{\partial^2u_2^\mu}{\partial x_3^2} =
\frac{1}{(\beta^\mu)^2}\frac{\partial^2u_2^\mu}{\partial t^2}\,,
\,\,\mu=u,d\,;
\end{equation*}
for conciseness, $\mu$ stands either for $u$\,, which corresponds to the layer, or for $d$\,, which corresponds to the halfspace.
Inserting expressions~(\ref{eq:LoveAnsatzU}) and (\ref{eq:LoveAnsatzD}) into the wave equation, we obtain
\begin{equation*}
\frac{\,{\rm d}^2A^\mu}{\,{\rm d}x_3^2}+\left(\frac{\omega^2}{(\beta^\mu)^2}-\kappa^2\right)A^\mu=0\,,\,\,\,\mu=u,d\,,
\end{equation*}
which is an ordinary differential equation for the amplitude of the $SH$ waves.
In view of physical requirements, we set the amplitude to decay in the halfspace, therefore we set $v_\ell <  \beta^d$\,, where $v_\ell=\omega/\kappa$ is the propagation speed of the guided wave. 
It also follows---from the dispersion relation of Love waves shown in expression~(\ref{eq:LoveDisp}), below---that $\beta^u<v_\ell $\,, since otherwise there is a sign incompatibility between the two sides of that expression.  
Thus,
\begin{equation*}
A^u =
C_1\exp\left(-\iota\sqrt{\frac{\omega^2}{(\beta^u)^2}-\kappa^2}\,x_3\right) + 
C_2\exp\left(\iota\sqrt{\frac{\omega^2}{(\beta^u)^2}-\kappa^2}\,x_3\right)\,,
\end{equation*}
\begin{equation*}
A^d=
C_4\exp\left(-\sqrt{\kappa^2-\frac{\omega^2}{(\beta^d)^2}}\,x_3\right)\,.
\end{equation*}

\noindent
For a notational convenience, we let
\begin{equation*}
s^u := \sqrt{\frac{v_\ell^2}{(\beta^u)^2}-1}\,,
\qquad 
s^d := \sqrt{1-\frac{v_\ell^2}{(\beta^d)^2}}\,,
\end{equation*}
so that we express the two general solutions as
\begin{equation}
\label{eq:LoveGenSols}
A^u = 
C_1\exp{\left(-\iota\kappa s^u x_3\right)}+C_2\exp{\left(\iota\kappa s^u x_3\right)}\,,
\qquad  
A^d = C_4\exp{\left(-\kappa s^d x_3\right)}\,.
\end{equation}
Thus, there is a sinusoidal dependence along the $x_3$-axis in the layer and an evanescent behaviour in the halfspace.
%%%%%%%%%%%%%%%%%%%%%%%%%%%%
\subsection{Boundary conditions}
%%%%%%%%%%%%%%%%%%%%%%%%%%%%
To find the specific solutions corresponding to expressions~(\ref{eq:LoveGenSols}), we need to find expressions for $C_1$\,, $C_2$ and $C_4$\,.
To do so, we need to consider the boundary conditions.
According to the continuity of stress expressed in terms of Hooke's law for an isotropic solid,
\begin{equation}
\label{eq:HookeLaw}
\sigma_{ij} =
\lambda\delta_{ij}\sum_{k=1}^3 \varepsilon_{kk} + 2\mu\varepsilon_{ij}
= (C_{11}-2C_{44})\,\delta_{ij}\sum_{k=1}^3\frac{\partial u_k}{\partial x_k} + 
C_{44}\left(\frac{\partial u_i}{\partial x_j}+\frac{\partial u_j}{\partial x_i}\right)\,,
\end{equation}
the boundary conditions are
\begin{equation}
\label{eq:LoveBC1}
C_{44}^u \frac{\partial u_2^u}{\partial x_3}=0\,,
\end{equation}
at the surface,~$x_3=0$\,, and
\begin{equation}
\label{eq:LoveBC2}
C_{44}^u \frac{\partial u_2^u}{\partial x_3}=
C_{44}^d \frac{\partial u_2^d}{\partial x_3}\,,
\end{equation}
\begin{equation}
\label{eq:LoveBC3}
u_2^u=u_2^d\,,
\end{equation}
at the interface,~$x_3=Z$\,.
Using expression~(\ref{eq:LoveAnsatzU}), we state expression~(\ref{eq:LoveBC1}), which is the boundary condition at the surface, as
\begin{equation}
\label{eq:LoveRedBC1}
-\iota\kappa s^u (C_1-C_2) =0\,.
\end{equation}
Using expressions~(\ref{eq:LoveAnsatzU}),  (\ref{eq:LoveAnsatzD}) and (\ref{eq:LoveRedBC1}) , we state expression~(\ref{eq:LoveBC2}), which is the boundary condition at the interface, as
\begin{equation}
\nonumber
C_{44}^u(-\iota\kappa s^u C_1 \,{\rm e}^{-\iota\kappa s^u Z})
+C_{44}^u(\iota\kappa s^u C_1 \,{\rm e}^{\iota\kappa s^u Z})=
C_{44}^d(-\kappa s^d C_4 \,{\rm e}^{-\kappa s^d Z})\,,
\end{equation}
which reduces to
\begin{equation}
\label{eq:LoveRedBC2}
C_1 (-2 C_{44}^u s^u\sin(\kappa s^u Z)) + 
C_4 ( C_{44}^ds^d \,{\rm e}^{-\kappa s^d Z})=0\,.
\end{equation}
Similarly, expression~(\ref{eq:LoveBC3}), which is the displacement at the interface, becomes
\begin{equation}
\label{eq:LoveRedBC3}
C_1 ( 2 \cos (\kappa s^u Z)) - C_4\,{\rm e}^{-\kappa s^d Z}=0\,.
\end{equation}
Equations (\ref{eq:LoveRedBC2}) and  (\ref{eq:LoveRedBC3})  form a linear system of equations for  $C_1$ and  $C_4$.
%%%%%%%%%%%%%%%%%%%%%%%%%%%%
\subsection{Dispersion relation}
%%%%%%%%%%%%%%%%%%%%%%%%%%%%
For nontrivial solutions for $C_1$ and $C_4$\,, the determinant of the coefficient matrix must be zero.
Explicitly, factoring out ${\rm e}^{-\kappa s^d Z}$ from the second column, we write
\begin{equation*}
\det[M_\ell]
=\det\left[
\begin{array}{cc}
-2 C_{44}^u s^u\sin(\kappa s^u Z)& C_{44}^ds^d \\
2 \cos (\kappa s^u Z) &  -1
\end{array}
\right]
=0\,.
\end{equation*}
Computing this determinant, we obtain
\begin{equation}
\label{eq:LoveDet1}
f(v_\ell):=2C_{44}^u s^u\sin(\kappa s^u Z)- 2C_{44}^ds^d \cos (\kappa s^u Z)=0 \,,
\end{equation}
which is the dispersion relation.
This equation, which is real, can be solved for $v_\ell$, given $Z$\,, $\omega$\,, $C_{44}^u$\,, $C_{44}^d$\,, $\rho^u$ and $\rho^d$\,.
Since the resulting expression for $v_\ell$ contains $\omega$\,, it means that Love waves are dispersive.

\begin{figure}
\begin{center}
\includegraphics[scale=0.5]{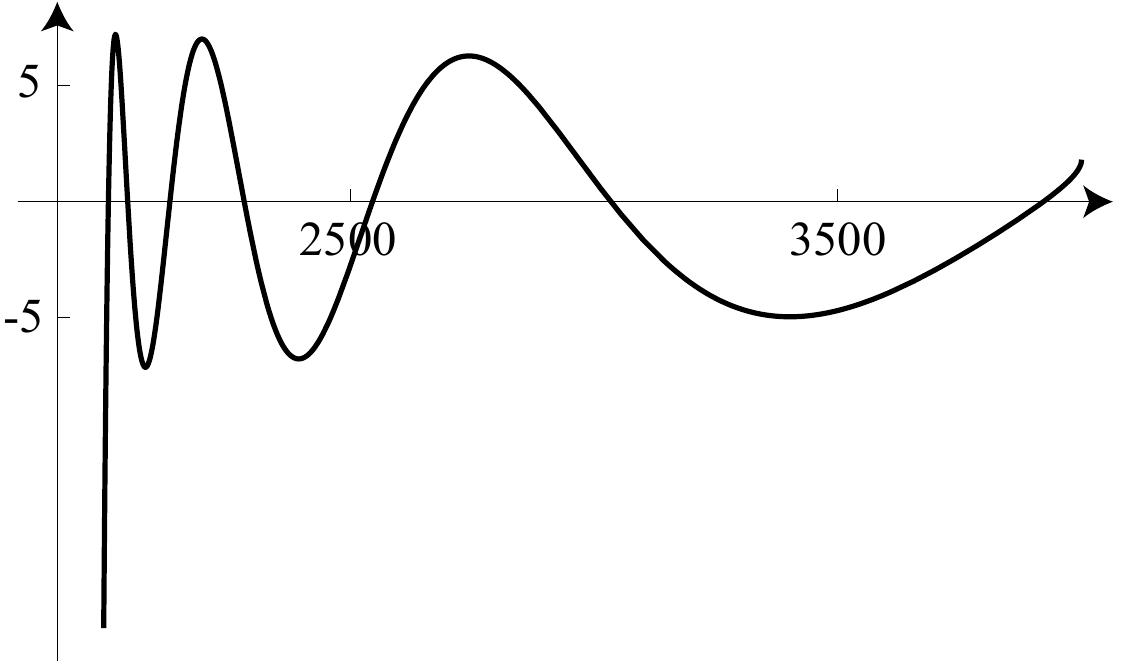}\hspace{1in}
\includegraphics[scale=0.5]{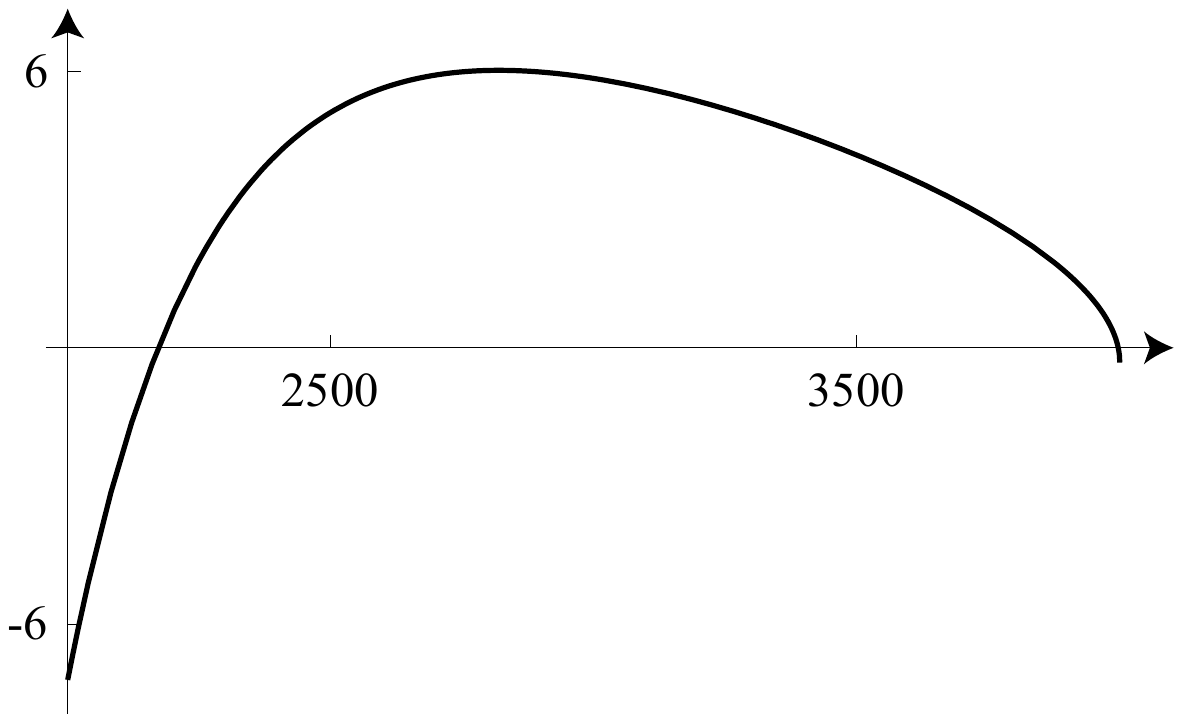}
%from love3x3.nb
\end{center}
\caption{\small{$\det[M_\ell]$ as a function of speed,~$v_\ell$\,.
On the left, for $\omega=90~{\rm s}^{-1}$\,, there are seven roots:
${v_\ell=2004.79~{\rm m}/{\rm s}}$\,,
$v_\ell=2044.33~{\rm m}/{\rm s}$\,, 
$v_\ell=2130.82~{\rm m}/{\rm s}$\,,
$v_\ell=2283.27~{\rm m}/{\rm s}$\,, 
$v_\ell=2546.14~{\rm m}/{\rm s}$\,,
$v_\ell=3035.03~{\rm m}/{\rm s}$ and 
$v_\ell=3921.62~{\rm m}/{\rm s}$\,.
On the right, for $\omega=15~{\rm s}^{-1}$\,,  
there are two roots:
$v_\ell=2172.48~{\rm m}/{\rm s}$ and $v_\ell=3997.01~{\rm m}/{\rm s}$\,.
The values on the vertical axes are to be multiplied by $10^{10}$\,.}}
\label{fig:love}
\end{figure}

For more derivation details, see~\citet[Chapter~6]{SlawinskiWS2016}, which in turn references, among others, \citet{Achenbach1973}, \citet{GrantWest1965} and \citet{Krebes2004}.
The dispersion relation for Love waves can be written as
\begin{equation}
\label{eq:LoveDisp}
\tan\left(\omega\sqrt{\frac{1}{(\beta^u)^2}-\frac{1}{v_\ell^2}}\,Z\right)
= 
\frac{
C_{44}^d\,\sqrt{\dfrac{\mathstrut 1}{v_\ell^2}-\dfrac{1}{(\beta^d)^2}}
}{
C_{44}^u\,\sqrt{\mathstrut\dfrac{\mathstrut 1}{(\beta^u)^2}-\dfrac{1}{v_\ell^2}}
}\,,
\end{equation}
which in our concise notation can be expressed as
\begin{equation*}
\tan(\kappa \,Z s^u)=\frac{C_{44}^d \,s^d}{C_{44}^u \,s^u}\,.
\end{equation*}
To examine the behaviour of the determinant, we use equation~(\ref{eq:LoveDet1}) for plotting purposes; explicitly, we write
\begin{equation*}
f(v_\ell) =
2\left(C_{44}^u\sqrt{\left(\frac{v_\ell}{\beta^u}\right)^{\!\!2}-1}\,\sin\left(\frac{\omega Z}{\beta^u}\sqrt{1-\left(\frac{\beta^u}{v_\ell}\right)^{\!\!2}}\,\right)- 
C_{44}^d\sqrt{1- \left(\frac{v_\ell}{\beta^d}\right)^{\!\!2}}\,\cos\left(\frac{\omega Z}{\beta^u}\sqrt{1-\left(\frac{\beta^u}{v_\ell}\right)^{\!\!2}}\,\right)\right)\,.
\end{equation*}
To plot the dispersion relation, we let the layer thickness $Z=500~{\rm m}$\,,  the  two elasticity parameters $C_{44}^u=0.88\times10^{10}~{\rm N}/{\rm m}^2$ and $C_{44}^d=4.16\times10^{10}~{\rm N}/{\rm m}^2$\,, and mass densities $\rho^u=2200~{\rm kg}/{\rm m}^3$ and $\rho^d=2600~{\rm kg}/{\rm m}^3$\,, then $\beta^u=2000~{\rm m}/{\rm s}$ and $\beta^d=4000~{\rm m}/{\rm s}$.
These parameters could exemplify a sandstone layer over a granite halfspace.

Examining the left and right plots of Figure~\ref{fig:love}, we see that, for high frequency,~$\omega=90~{\rm s}^{-1}$, the speed of the fundamental mode of the Love wave,~$v_\ell=2004.79~{\rm m}/{\rm s}$\,, approaches $\beta^u$, from above.
The values of $v_\ell$ are computed numerically.
This result is in agreement with~\citet[p.~196]{Udias1999}, and with the general theory of Love waves.
%%%%%%%%%%%%%%%%%%%%%%%%%%%%
\subsection{Sensitivity of dispersion relation}
%%%%%%%%%%%%%%%%%%%%%%%%%%%%
We wish to examine effects of various parameters on the dispersion relation.
To do so, we examine their effect on the value of the determinant,~$\det[M_\ell]$\,.
In this case, we do not examine the determinant as a function of $v_\ell$ but as a function of $C_{44}^d$ and $C_{44}^u$ for a fixed $v_\ell$\,.

\begin{figure}
\begin{center}
\includegraphics[scale=0.5]{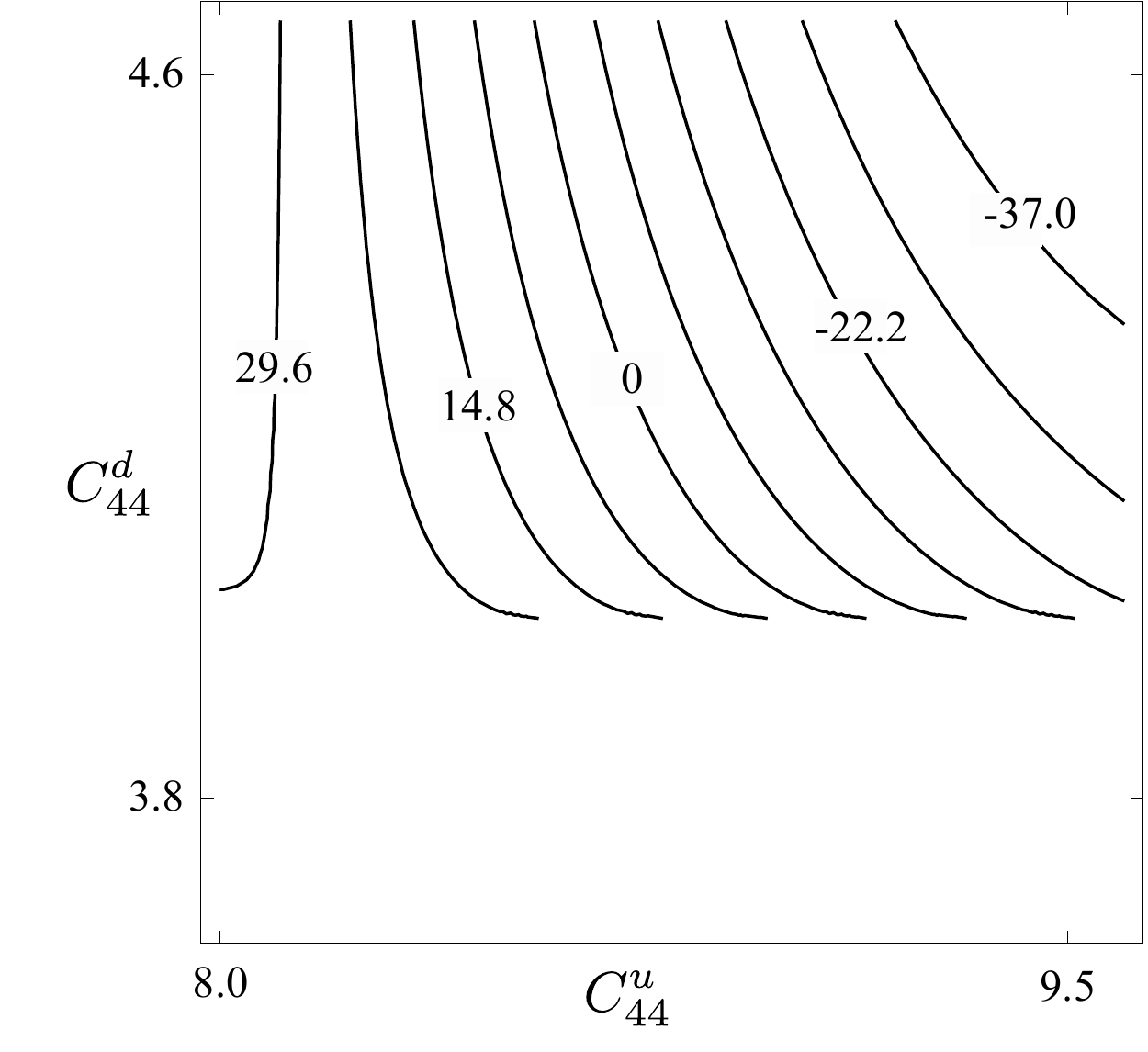}\hspace*{1in}\includegraphics[scale=0.5]{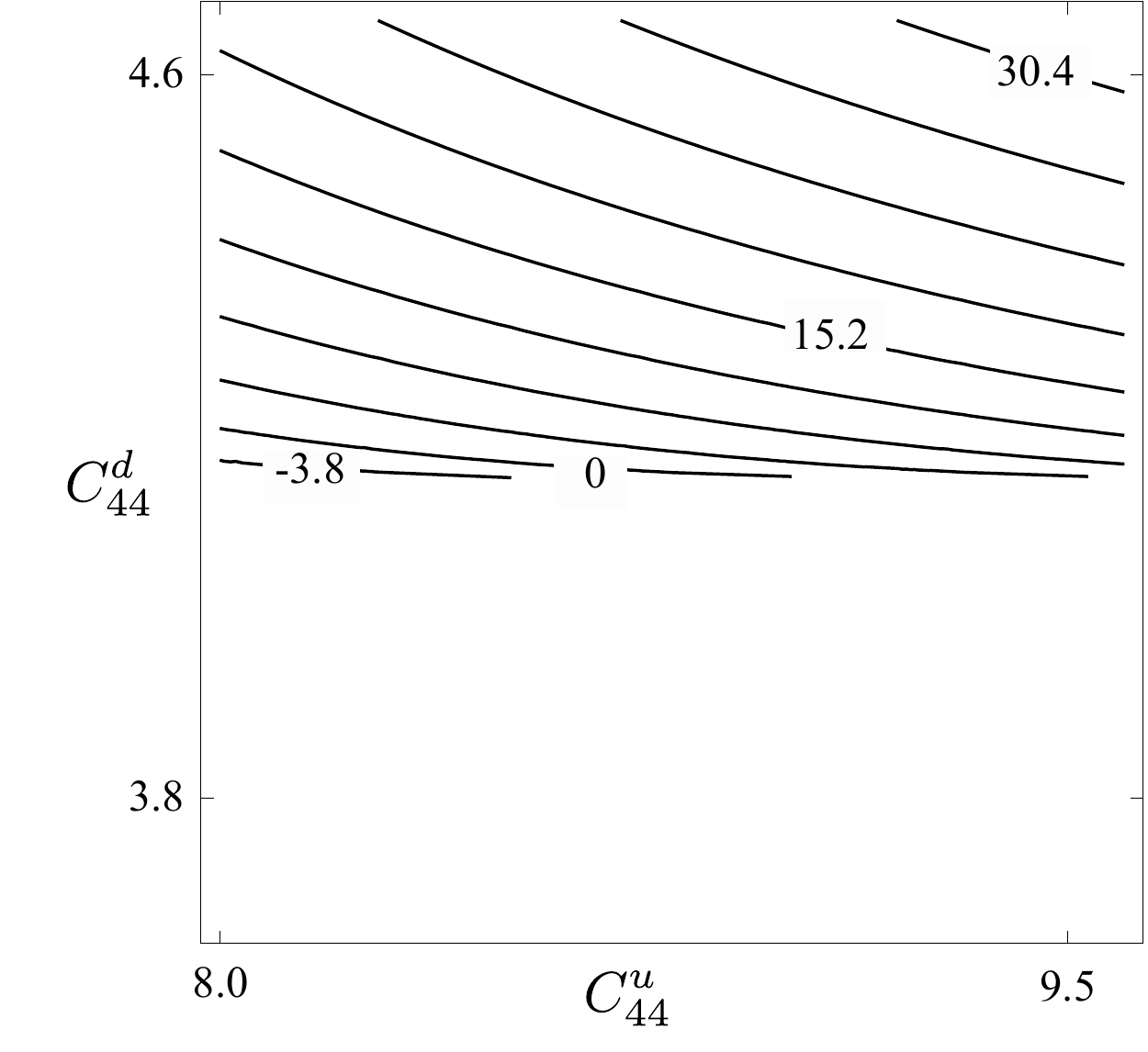}
%from love3x3contour-2.nb
\end{center}
\caption{\small{
$\det[M_\ell]/10^{9}$ as a function of the elasticity parameters, $C_{44}^u$ and $C_{44}^d$\,, 
for
$\omega=90~{\rm s}^{-1}$ and 
$v_\ell=3921.62~{\rm km}/{\rm s}$\,, on the left,
and for
$\omega=15~{\rm s}^{-1}$ and 
$v_\ell=3997.01~{\rm km}/{\rm s}$\,, on the right.
The values on the horizontal and vertical axes are to be multiplied by $10^9$ and $10^{10}$\,, respectively.}}
\label{fig:lc-vhigh}
\end{figure}

\begin{figure}
\begin{center}
\includegraphics[scale=0.5]{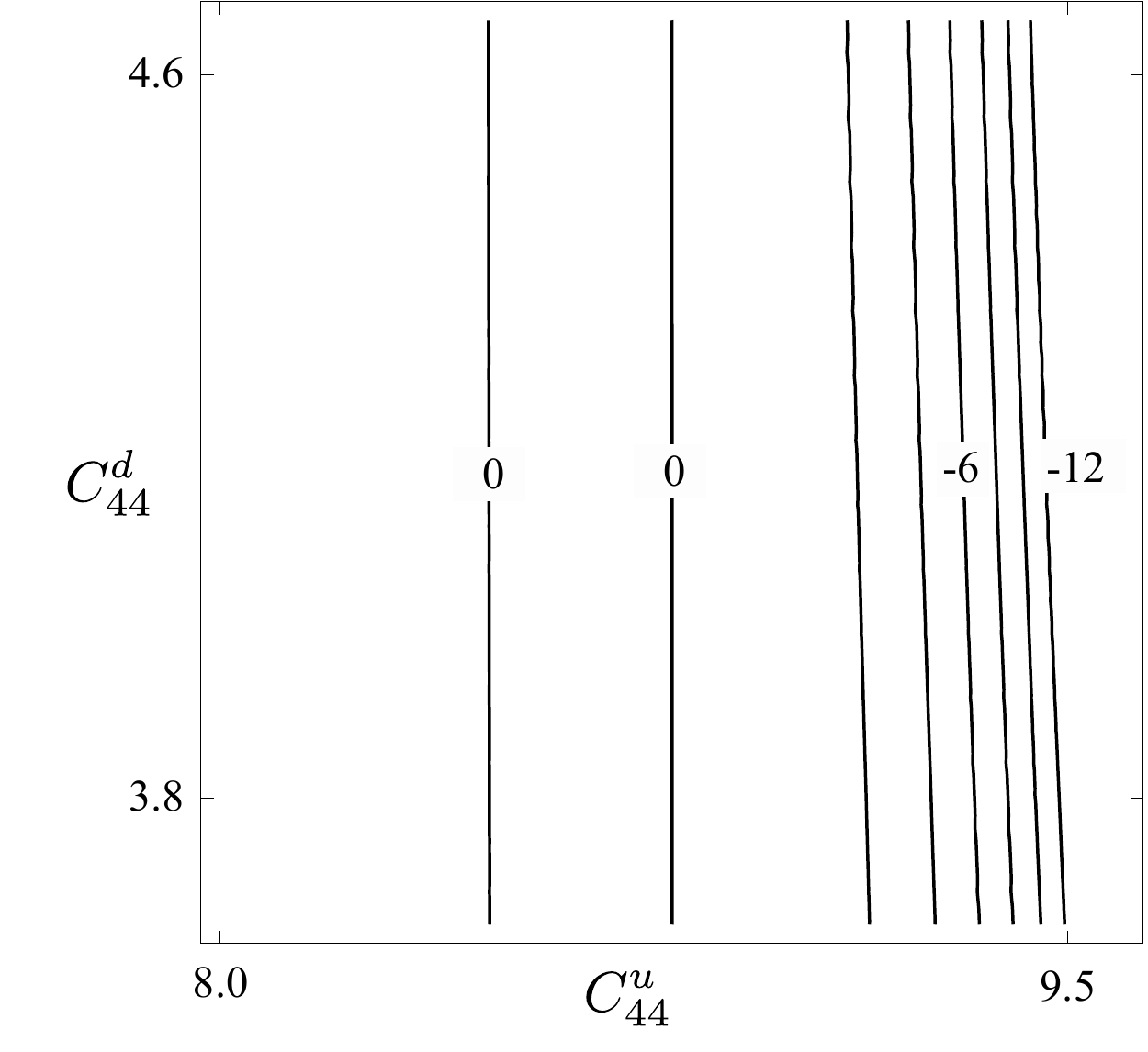}\hspace*{1in}\includegraphics[scale=0.5]{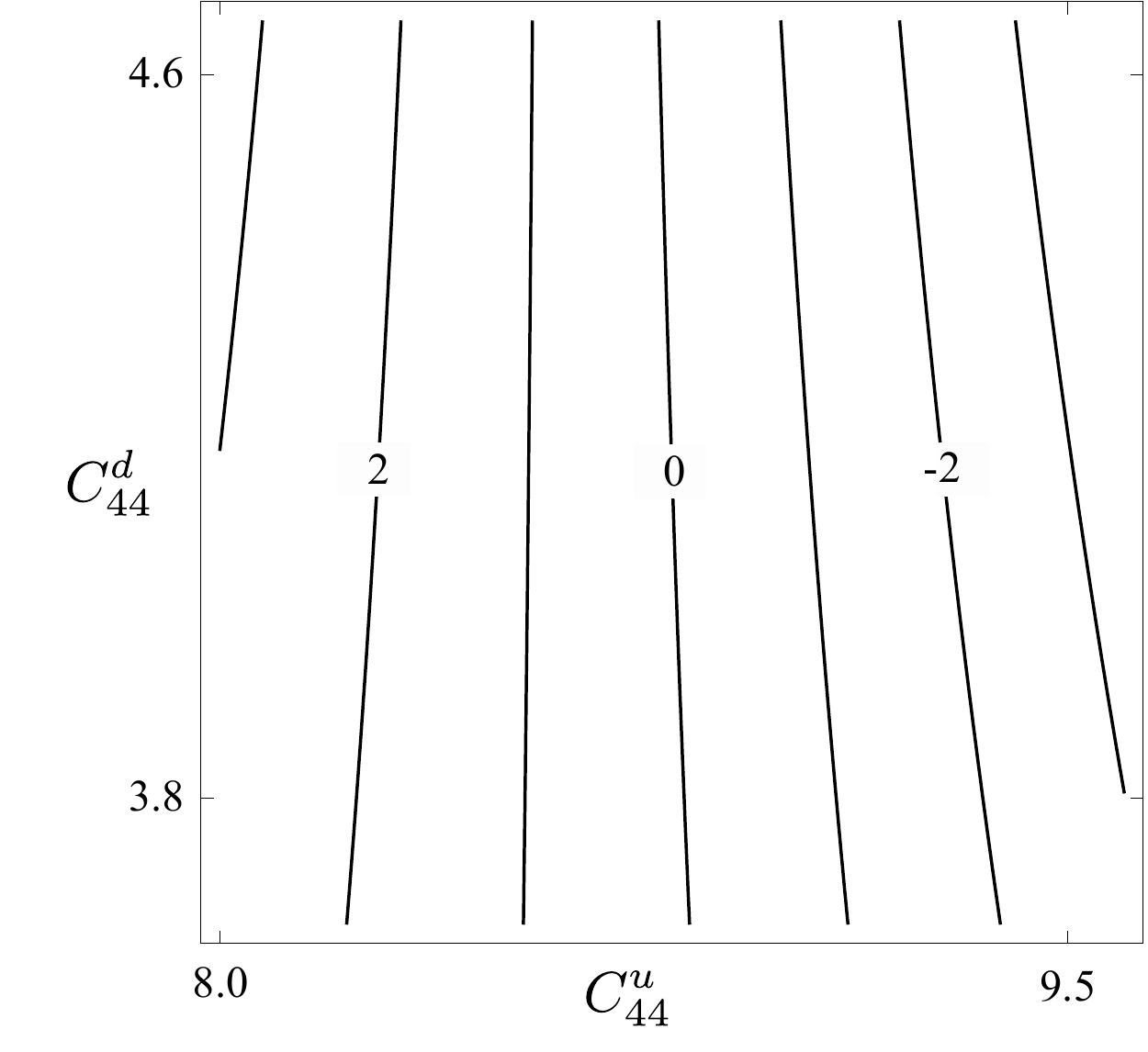}
%from love3x3contour-2.nb
\end{center}
\caption{\small{
$\det[M_\ell]/10^{12}$\,, on the left, and $\det[M_\ell]/10^{10}$\,, on the right, as a function of the elasticity parameters, $C_{44}^u$ and $C_{44}^d$\,, where
$\omega=90~{\rm s}^{-1}$ and
$v_\ell=2004.79~{\rm km}/{\rm s}$\,,
and where
$\omega=15~{\rm s}^{-1}$ and
$v_\ell=2172.48~{\rm km}/{\rm s}$\,,
for left and right, respectively.
The values on the horizontal and vertical axes are to be multiplied by $10^9$ and $10^{10}$\,, respectively.}}
\label{fig:lc-vlow}
\end{figure}

\begin{figure}
\begin{center}
\includegraphics[scale=0.5]{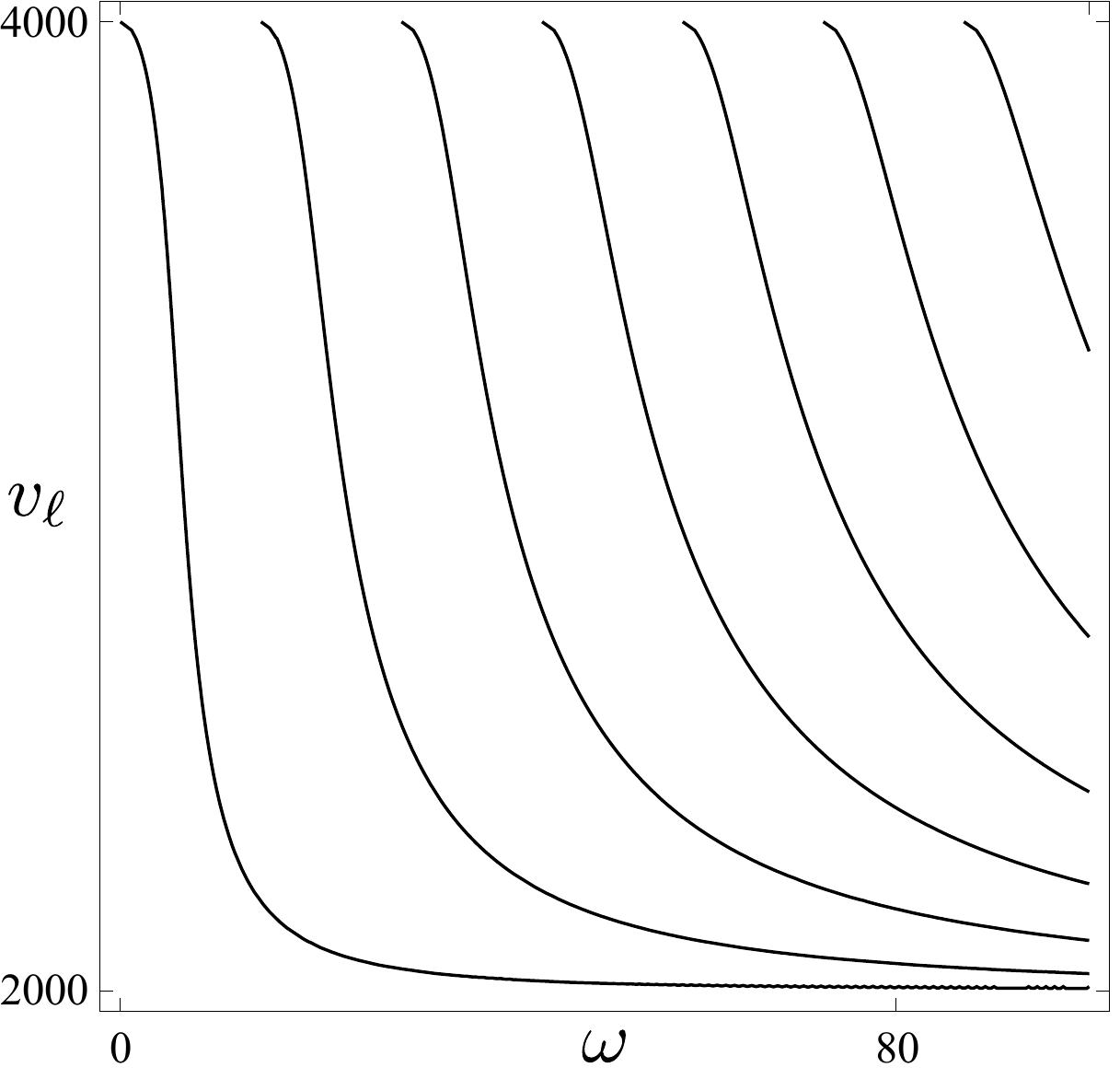}
\end{center}
\caption{\small{Zero lines of the Love-wave dispersion curves,~$\det[M_\ell]=0$\,, as a function of speed,~$v_\ell$\,, and frequency,~$\omega$\,.}}
%from love3x3contour-w5.nb
\label{fig:love3x3contourw5-2}
\end{figure}

We examine the seventh root of the left plot of Figure~\ref{fig:love}, which---calculated numerically---is $v_\ell=3922~{\rm m}/{\rm s}$\,, and the second root of the right plot of 
Figure~\ref{fig:love}, which is $v_\ell=3997~{\rm m}/{\rm s}$.
The left and right plots of Figure~\ref{fig:lc-vhigh} are the corresponding contour plots of the values of $\det[M_\ell]/10^{9}$ with varying $C_{44}^u$ and $C_{44}^d$\,. 
In both cases,  $\det[M_\ell]$ is sensitive to variations in both $C_{44}^u$ and $C_{44}^d$\,. 
However, if we examine the first root of the left plot of Figure~\ref{fig:love}, $v_\ell=2005~{\rm m}/{\rm s}$\,, and the first root of the right plot of Figure~\ref{fig:love}, $v_\ell=2172~{\rm m}/{\rm s}$\,, in Figure~\ref{fig:lc-vlow} there are near vertical lines at $C_{44}^u=0.88\times10^{10}$ indicating a sensitivity to $C_{44}^u$ but not to $C_{44}^d$.
This is more pronounced for the left plot.
This indicates that for a given wavelength, a solution with speed closer to $\beta^u$ is less sensitive to $C_{44}^d$ than a solution with greater speed. 
Also, the right plot of Figure~\ref{fig:lc-vhigh} indicates a greater sensitivity to $C_{44}^d$ for the lower frequency, which is tantamount to longer wavelength.
Similar linear patterns of ridges and valleys are obtained for other combinations of parameters, such as $\beta^u$ and~$Z$. 
 
Such a behaviour is to be expected.
We can write expression~(\ref{eq:LoveAnsatzU}), which corresponds to the nonzero component of the displacement, together with expression~(\ref{eq:LoveGenSols}), which contains the general solution for the $SH$ waves in the layer, as
\begin{equation*}
u_2^u(x_1,x_3,t) =
C_1\exp{\left(-\iota\kappa s^u x_3\right)} \exp[\iota(\kappa x_1-\omega t)] + C_2\exp{\left(\iota\kappa s^u x_3\right)}\exp[\iota(\kappa x_1-\omega t)]\,.
\end{equation*}
This can be interpreted as a superposition of two $SH$ waves within the elastic layer.
Both waves travel obliquely with respect to the surface and interface; one wave travels upwards, the other downwards.
Their wave vectors are $\displaystyle \overline{k}_{\pm}:=(\kappa, 0, \pm \kappa s^u)\,.$ 
Thus---since $|\overline{k}_\pm|=\sqrt{\kappa^2+(\kappa\,s^u)^2}\,$---we have $\displaystyle |\overline{k}_\pm|=\kappa\sqrt{1+(s^u)^2}$\,, which is $\kappa\,v_\ell/\beta^u$\,, where  $\beta^u/v_\ell=\kappa/|k_\pm|=\sin \theta$\,, with $\theta $ standing for the angle between $\overline{k}_{\pm}$ and the $x_3$-axis.
Thus, $\theta$ is the angle between the $x_3$-axis and a wavefront normal; therefore, it is the propagation direction of a wavefront.
Hence, for the case where $v_\ell$ is only slightly larger than $\beta^u$\,, it follows that both $SH$ waves propagate nearly horizontally; in other words, their propagation directions are nearly parallel to the interface.
In such a case, the resulting Love wave is less sensitive to the material properties below the interface than for the case of $\beta^u/v_\ell\ll 1$\,.

Depending on the propagation direction,~$\theta$\,, of the $SH$ waves, there is a distinction in the sensitivity to the material properties of the halfspace.
This distinction is more pronounced for short wavelengths, in comparison with the layer thickness,~$Z$\,.
For any angle, the longer the wavelength the more sensitivity of the wave to material properties below the interface, and thus the distinction---as a function of the propagation direction---is diminished.

Also, the superposition helps us understand the requirement of $\beta^u<v_\ell<\beta^d$\,.
The lower limit, which we can write as $\sin \theta=\beta^u/\beta^u=1$\,, corresponds to waves that propagate along the $x_1$-axis, and hence, do not exhibit any interference associated with the presence of the horizontal surface or interface.
Formally, the lower limit is required by the real sine function.

The upper limit is introduced to ensure an exponential amplitude decay in the halfspace.
Herein, we can write the upper limit as $\sin \theta=\beta^u/\beta^d$\,, which---in general---implies the $SH$ waves propagate obliquely. 
In the extreme case, if $\beta^d\to\infty$\,, the waves propagate vertically.
Also, this case is tantamount to total internal reflection, since it corresponds to a rigid halfspace, $C_{44}^d\to\infty$\,; such a case is discussed by~\mbox{\citet[Section~6.3.1]{SlawinskiWS2016}}.
%%%%%%%%%%%%%%%%%%%%%%%%%%%%
\section{Quasi-Rayleigh waves}
\label{sec:qRwaves}
%%%%%%%%%%%%%%%%%%%%%%%%%%%%
\subsection{Material properties and wave equations}
%%%%%%%%%%%%%%%%%%%%%%%%%%%%
\footnote{The formulation in this section is similar to that of~\citet[Section~10.4]{Udias1999} except that we set $x_3$ to be positive downwards with the free surface at $x_3=0$ and interface at $x_3=Z$\,, whereas~\citet{Udias1999} sets $x_3$ to be positive upwards with the free surface at $x_3=H$ and interface at $x_3=0$\,.
More importantly, unlike~\citet{Udias1999}, we do not restrict Poisson's ratio in the layer and in the halfspace to~$1/4$\,.}
To consider a quasi-Rayleigh wave within the model of material properties discussed in Section~\ref{sec:LovWav}, we also need to specify $C_{11}^u$ and $C_{11}^d$\,, which are elasticity parameters  in the layer and the halfspace.
Hence, the corresponding $P$-wave propagation speeds are $\alpha^u=\sqrt{C_{11}^u\strut/\rho^u}$ and $\alpha^d=\sqrt{C_{11}^d/\rho^d}$.

Using the Helmholtz decomposition theorem, we express the displacements as
\begin{equation}
\label{eq:QRpotentials}
u^\mu=\nabla{\mathcal P}^\mu+\nabla\times\mbox{\boldmath${\mathcal S}$}^\mu\,,\quad u_1^\mu=\frac{\partial {\mathcal P}^\mu}{\partial x_1}
-\frac{\partial {\mathcal S}^\mu}{\partial x_3}\,,\quad u_3^\mu=\frac{\partial {\mathcal P}^\mu}{\partial x_3}+
\frac{\partial {\mathcal S}^\mu}{\partial x_1}\,,\quad u_2^\mu=0\,,\,\,\mu=u,d,
\end{equation}
where we use the gauge condition outlined in~\citet[Section~6.2]{SlawinskiWS2016} and set ${\mathcal S}_2^\mu={\mathcal S}^\mu$, for brevity.
$\mathcal P^\mu$ denotes the scalar potential and $\mbox{\boldmath${\mathcal S}$}^\mu=[{\mathcal S}_1^\mu,{\mathcal S}_2^\mu,{\mathcal S}_3^\mu]$ denotes the vector potential, which herein is $\mbox{\boldmath${\mathcal S}$}^\mu=[0,{\mathcal S}^\mu,0]$\,.
Potentials allow us to consider the coupling between the $P$ and $SV$ waves.
The pertinent wave equations are
\begin{equation}
\label{eq:QRWaveEqns}
\nabla^2 {\mathcal P}^\mu-\frac{1}{(\alpha^\mu)^2}\frac{\partial^2 {\mathcal P}^\mu}{\partial t^2}=0\,\,,\,\,\,
\nabla^2 {\mathcal S}^\mu-\frac{1}{(\beta^\mu)^2}\frac{\partial^2 {\mathcal S}^\mu}{\partial t^2}=0\,,\,\,\,\mu=u,d,
\end{equation}
which correspond to the $P$ waves and $SV$ waves, respectively.
%%%%%%%%%%%%%%%%%%%%%%%%%%%%
\subsection{Solutions of wave equations}
%%%%%%%%%%%%%%%%%%%%%%%%%%%%
Let the trial solutions for the corresponding wave equations be
\begin{equation}
\label{eq:QRAnsatzP}
{\mathcal P}^\mu=A^\mu(x_3)\exp(\iota(\kappa x_1-\omega t))\,,\quad\mu=u,d\,,
\end{equation}
\begin{equation}
\label{eq:QRAnsatzS}
{\mathcal S}^\mu=B^\mu(x_3)\exp(\iota(\kappa x_1-\omega t))\,,\quad\mu=u,d\,.
\end{equation}
Inserting solutions~(\ref{eq:QRAnsatzP}) and (\ref{eq:QRAnsatzS}) into equations~(\ref{eq:QRWaveEqns}) leads to
\begin{equation*}
\frac{\,{\rm d}^2 A^\mu}{\,{\rm d}x_3^2}+\left(\frac{\omega^2}{(\alpha^\mu)^2}-\kappa^2\right)A^\mu=0\,,\,\,\,
\frac{\,{\rm d}^2 B^\mu}{\,{\rm d}x_3^2}+\left(\frac{\omega^2}{(\beta^\mu)^2}-\kappa^2\right)B^\mu=0\,,\,\,\,\mu=u,d\,,
\end{equation*}
which are ordinary differential equations for amplitudes $A^\mu$ and $B^\mu$. 
Similarly to the derivation of Love waves in Section~\ref{sub:DerSH}, we require displacements to decay  within the halfspace.
Expressions~(\ref{eq:QRpotentials}), which denote displacements, entail
\begin{equation*}
\kappa^2-\omega^2/(\alpha^d)^2 > 0\quad{\rm and}\quad\kappa^2-\omega^2/(\beta^d)^2 > 0\,.
\end{equation*}
Thus, we obtain four general solutions, which we write as
\begin{equation}
A^u =
C_1\exp\left(-\iota\sqrt{\frac{\omega^2}{(\alpha^u)^2}-\kappa^2}\,x_3\right) + 
C_2\exp\left(\iota\sqrt{\frac{\omega^2}{(\alpha^u)^2}-\kappa^2}\,x_3\right)\,,
\label{eq:QRGenSolAu}
\end{equation}
\begin{equation*}
A^d =
C_4\exp\left(-\sqrt{\kappa^2-\frac{\omega^2}{(\alpha^d)^2}}\,x_3\right)\,,
\end{equation*}
\begin{equation*}
B^u =
D_1\exp\left(-\iota\sqrt{\frac{\omega^2}{(\beta^u)^2}-\kappa^2}\,x_3\right) + 
D_2\exp\left(\iota\sqrt{\frac{\omega^2}{(\beta^u)^2}-\kappa^2}\,x_3\right)\,,
\end{equation*}
\begin{equation*}
B^d =
D_4\exp\left(-\sqrt{\kappa^2-\frac{\omega^2}{(\beta^d)^2}}\,x_3\right)\,.
\end{equation*}
As in Section~\ref{sub:DerSH}, we invoke $\kappa=\omega/v_{r}$\,, where $v_{r}$ is the propagation speed.
Herein, this speed corresponds to the quasi-Rayleigh wave.
For a notational convenience, we let
\begin{equation*}
r^u:=\sqrt{\frac{v_{r}^2}{(\alpha^u)^2}-1}\,,\qquad
s^u:=\sqrt{\frac{v_{r}^2}{(\beta^u)^2}-1}\,,\qquad
r^d:=\sqrt{1-\frac{v_{r}^2}{(\alpha^d)^2}}\,,\qquad
s^d:=\sqrt{1-\frac{v_{r}^2}{(\beta^d)^2}}\,.
\end{equation*}
Our assumption about the behaviour of solutions in the halfspace forces $r^d$ and $s^d$ to be real. Thus, we write the nonzero components of the displacement vector as
\begin{align}
\label{eq:QRuD1}
u_1^d=&\frac{\partial{\mathcal P}^d}{\partial x_1}-\frac{\partial {\mathcal S}^d}{\partial x_3}\\
\nonumber=&\left[\iota\kappa C_4\exp\left(-\kappa r^d x_3\right)
+D_4\kappa s^d\exp\left(-\kappa s^d x_3\right)\right]
\exp(\iota(\kappa x_1-\omega t))\,,
\end{align}
\begin{align}
u_3^d=&\frac{\partial{\mathcal P}^d}{\partial x_3}+\frac{\partial{\mathcal S}^d}{\partial x_1}\\
\nonumber=&\left[-C_4\kappa r^d\exp\left(-\kappa r^d x_3\right)
+\iota\kappa D_4\exp\left(-\kappa s^d x_3\right)\right]
\exp(\iota(\kappa x_1-\omega t))\,,
\end{align}
\begin{align}
\label{eq:QRuU1}
u_1^u=&\frac{\partial{\mathcal P}^u}{\partial x_1}-\frac{\partial{\mathcal S}^u}{\partial x_3}\\
\nonumber=&\left[\iota\kappa C_1 \exp(-\iota\kappa r^u x_3)+\iota\kappa C_2\exp(\iota\kappa r^u x_3)\right.\\
\nonumber &\left. \qquad+\,\iota\kappa s^u D_1\exp(-\iota\kappa s^u x_3)-\iota\kappa s^u D_2\exp(\iota\kappa s^u x_3)
\right]
\exp(\iota(\kappa x_1-\omega t))\,,
\end{align}
\begin{align}
\label{eq:QRuU3}
u_3^u=&\frac{\partial{\mathcal P}^u}{\partial x_3}+\frac{\partial{\mathcal S}^u}{\partial x_1}\\
\nonumber=&\left[-\iota\kappa r^u C_1 \exp(-\iota\kappa r^u x_3)+\iota\kappa r^u C_2\exp(\iota\kappa r^u x_3)\right.\\
\nonumber &\left. \qquad+\,\iota\kappa D_1\exp(-\iota\kappa s^u x_3)+\iota\kappa D_2\exp(\iota\kappa s^u x_3)
\right]
\exp(\iota(\kappa x_1-\omega t))\,,
\end{align}
which allows us to apply the boundary conditions.
%%%%%%%%%%%%%%%%%%%%%%%%%%%%
\subsection{Boundary conditions}
%%%%%%%%%%%%%%%%%%%%%%%%%%%%
Let us examine expressions~(\ref{eq:QRuD1}) to  (\ref{eq:QRuU3}), in view of Hooke's law, which is stated in expression~(\ref{eq:HookeLaw}), and of the following boundary conditions.
At $x_3=0$\,, the conditions are $\sigma_{33}^u=\sigma_{31}^u=0$\,; hence, the first condition implies
\begin{equation*}
\left.\sigma_{31}^u\right|_{x_3=0}=0\Rightarrow \left.\frac{\partial u_1^u}{\partial x_3}\right|_{x_3=0}= -\left.\frac{\partial u_3^u}{\partial x_1}\right|_{x_3=0}\,.
\end{equation*}
Factoring out $\exp(\iota(\kappa x_1-\omega t))$\,, we write
\begin{equation*}
\kappa^2 r^u C_1-\kappa^2 r^u C_2+\kappa^2 (s^u)^2D_1 +\kappa^2 (s^u)^2D_2
=-\kappa^2 r^u C_1+\kappa^2 r^u C_2+\kappa^2 D_1 +\kappa^2 D_2\,,
\end{equation*}
which, upon rearranging and factoring out $\kappa^2$, we rewrite as
\begin{equation}
\label{eq:QRBC1}
2r^u(C_1-C_2)+[(s^u)^2-1](D_1+D_2)=0\,.
\end{equation}
The second condition implies
\begin{equation*}
\left.\sigma_{33}^u\right|_{x_3=0}=0\Rightarrow \left.\left[(C_{11}^u-2C_{44}^u)
\left(\frac{\partial u_1^u}{\partial x_1}+\frac{\partial u_3^u}{\partial x_3}\right)
+2C_{44}^u\left(\frac{\partial u_3^u}{\partial x_3}\right)\right]\right|_{x_3=0}=0\,,
\end{equation*}
which can be rearranged to
\begin{equation*}
 \left.\left[(C_{11}^u-2C_{44}^u)
\left(\frac{\partial u_1^u}{\partial x_1}\right)
+C_{11}^u\left(\frac{\partial u_3^u}{\partial x_3}\right)\right]\right|_{x_3=0}=0\,,
\end{equation*}
and, upon factoring out $\kappa^2\exp(\iota(\kappa x_1-\omega t))$\,, further reduces to
\begin{equation}
(C_{11}^u-2C_{44}^u)\left[-(C_1+C_2)-s^u(D_1-D_2)\right]
+C_{11}^u\left[-(r^u)^2(C_1+C_2)+s^u(D_1-D_2)\right]=0\,.
\label{eq:sig330}
\end{equation}
At $x_3=Z$\,, the boundary conditions are
\begin{equation*}
u_1^u|_{x_3=Z}=u_1^d|_{x_3=Z}\,,\quad u_3^u|_{x_3=Z}=u_3^d|_{x_3=Z}\,,\quad
\sigma_{33}^u|_{x_3=Z}=\sigma_{33}^d|_{x_3=Z}\,,\quad
\sigma_{31}^u|_{x_3=Z}=\sigma_{31}^d|_{x_3=Z}
\,.
\end{equation*}
Factoring out  $\kappa \exp(\iota(\kappa x_1-\omega t))$\,, the first condition becomes
\begin{align}
\iota C_1\exp(-\iota\kappa r^u Z)+\iota C_2\exp(\iota\kappa r^u Z) + \iota s^u D_1
& \exp(-\iota\kappa s^u Z)-\iota s^u D_2\exp(\iota\kappa s^u Z)\nonumber\\
=&\,\iota C_4\exp(-\kappa r^d Z)+s^d D_4\exp(-\kappa s^d Z)\,.
\label{eq:u1Z}
\end{align}
Similarly,  the second condition implies
\begin{align}
-\iota r^u C_1\exp(-\iota\kappa r^u Z)+\iota r^u C_2\exp(\iota\kappa r^u Z) + \iota 
&D_1 \exp(-\iota\kappa s^u Z)+\iota D_2\exp(\iota\kappa s^u Z)\nonumber\\
=&-r^d C_4\exp(-\kappa r^d Z)+\iota D_4\exp(-\kappa s^d Z)\,.
\label{eq:u3Z}
\end{align}
The third condition,
\begin{equation*}
 \left.\left[(C_{11}^u-2C_{44}^u)
\left(\frac{\partial u_1^u}{\partial x_1}\right)
+C_{11}^u\left(\frac{\partial u_3^u}{\partial x_3}\right)\right]\right|_{x_3=Z}
=
 \left.\left[(C_{11}^d-2C_{44}^d)
\left(\frac{\partial u_1^d}{\partial x_1}\right)
+C_{11}^d\left(\frac{\partial u_3^d}{\partial x_3}\right)\right]\right|_{x_3=Z}\,,
\end{equation*}
upon factoring out $\kappa^2\exp(\iota(\kappa x_1-\omega t))$\,, becomes
{\small
\begin{align}
(C_{11}^u&-2C_{44}^u)\left[-C_1\exp(-\iota\kappa r^u Z)
-C_2\exp(\iota\kappa r^u Z)
-s^u D_1\exp(-\iota\kappa s^u Z)
+s^u D_2\exp(\iota\kappa s^u Z)\right]\nonumber\\
+\,&C_{11}^u\left[-(r^u)^2 C_1\exp(-\iota\kappa r^u Z)
-(r^u)^2C_2\exp(\iota\kappa r^u Z)
+s^u D_1 \exp(-\iota\kappa s^u Z)
-s^u D_2\exp(\iota\kappa s^u Z)\right]\nonumber\\
&=
(C_{11}^d-2C_{44}^d)\left[
-C_4\exp(-\kappa r^d Z)
+\iota s^d D_4\exp(-\kappa s^d Z)\right]\nonumber\\
&\qquad+C_{11}^d\left[
(r^d)^2 C_4 \exp(-\kappa r^d Z)
-\iota s^d D_4 \exp(-\kappa s^d Z)\right]\,.
\label{eq:sig33Z}
\end{align}}
The fourth condition,
\begin{equation*}
\left. C_{44}^u\left(\frac{\partial u_1^u}{\partial x_3}+\frac{\partial u_3^u}{\partial x_1}\right)
\right|_{x_3=Z}=
\left. C_{44}^d\left(\frac{\partial u_1^d}{\partial x_3}+\frac{\partial u_3^d}{\partial x_1}\right)
\right|_{x_3=Z}\,,
\end{equation*}
implies
{\small
\begin{align}
C_{44}^u&\left[
r^u C_1\exp(-\iota\kappa r^u Z)
-r^u C_2\exp(\iota\kappa r^u Z)
+(s^u)^2 D_1 \exp(-\iota\kappa s^u Z)
+(s^u)^2 D_2 \exp(\iota\kappa s^u Z)\right.\nonumber\\
&\left.+\,r^u C_1\exp(-\iota\kappa r^u Z)
-r^u C_2\exp(\iota\kappa r^u Z)
-D_1\exp(-\iota\kappa s^u Z)
-D_2\exp(\iota\kappa s^u Z)\right]\nonumber\\
&\quad =
C_{44}^d\left[
-\iota r^d C_4\exp(-\kappa r^d Z)
-(s^d)^2D_4\exp(-\kappa s^d Z)
-\iota r^d C_4\exp(-\kappa r^d Z)
-D_4\exp(-\kappa s^d Z)\right]\,.
\label{eq:QRBC6}
\end{align}}
For a notational convenience, we let 
\begin{equation}
\label{eq:QRConvI}
a':=\kappa r^u Z\,, 
a=\kappa r^d Z\,, 
C_1':=C_1\,{\rm e}^{-\iota a'}\,,
C_2':=C_2\,{\rm e}^{\iota a'}\,, 
C_4':=C_4\,{\rm e}^{-a}\,,
\end{equation}
\begin{equation}
\label{eq:QRConvII}
b':=\kappa s^u Z\,,
b=\kappa s^d Z\,,
D_1':=D_1\,{\rm e}^{-\iota b'}\,,
D_2':=D_2\,{\rm e}^{\iota b'}\,,
D_4':=D_4\,{\rm e}^{-b}\,.
\end{equation}
Thus, conditions~(\ref{eq:QRBC1}) to~(\ref{eq:QRBC6}) can be written as
\begin{equation}
\label{eq:QRModBC1}
2r^u\,{\rm e}^{\iota a'}C_1'-2r^u\,{\rm e}^{-\iota a'}C_2'+[(s^u)^2-1]\,{\rm e}^{\iota b'}D_1'
+[(s^u)^2-1]\,{\rm e}^{-\iota b'}D_2'=0\,,
\end{equation}
\begin{align}
\label{eq:QRModBC2}
[-(C_{11}^u-&2C_{44}^u)-C_{11}^u (r^u)^2]\,{\rm e}^{\iota a'}C_1'
+[-(C_{11}^u-2C_{44}^u)-C_{11}^u (r^u)^2]\,{\rm e}^{-\iota a'}C_2'\nonumber\\
&+2C_{44}^u s^u\,{\rm e}^{\iota b'}D_1'-2C_{44}^u s^u\,{\rm e}^{-\iota b'}D_2'=0\,,
\end{align}
\begin{equation}
\label{eq:QRModBC3}
\iota C_1'+\iota C_2' +\iota s^u D_1' -\iota s^u D_2'-\iota C_4'-s^d D_4'=0\,,
\end{equation}
\begin{equation}
-\iota r^u C_1'+\iota r^u C_2'+\iota D_1'+\iota D_2'+r^d C_4'-\iota D_4'=0\,,
\label{eq:QRModBC4}
\end{equation}
\begin{align}
\label{eq:QRModBC5temp}
-(C_{11}^u&-2C_{44}^u)C_1'-(C_{11}^u-2C_{44}^u)C_2'-(C_{11}^u-2C_{44}^u)s^u D_1'
+(C_{11}^u-2C_{44}^u)s^u D_2'\nonumber\\
&-C_{11}^u (r^u)^2C_1'-C_{11}^u(r^u)^2C_2'+C_{11}^u s^u D_1'-C_{11}^u s^u D_2'
+(C_{11}^d-2C_{44}^d)C_4'
\nonumber\\&\qquad-\iota(C_{11}^d-2C_{44}^d)s^d D_4'
-C_{11}^d (r^d)^2 C_4'+\iota C_{11}^d s^d D_4'=0\,,
\end{align}
\begin{align}
\label{eq:QRModBC6temp}
C_{44}^u& r^u C_1' - C_{44}^u r^u C_2' +C_{44}^u (s^u)^2D_1'+C_{44}^u (s^u)^2 D_2'
+C_{44}^u r^u C_1'-C_{44}^u r^u C_2' -C_{44}^u D_1' -C_{44}^u D_2'\nonumber\\
&+\iota r^d C_{44}^d C_4'+C_{44}^d (s^d)^2 D_4'+\iota r^d C_{44}^d C_4'+C_{44}^d D_4'=0\,.
\end{align}
Simplifying, we write conditions~(\ref{eq:QRModBC5temp}) and (\ref{eq:QRModBC6temp}) as
\begin{align}
\label{eq:QRModBC5}
-[C_{11}^u&-2C_{44}^u+(r^u)^2C_{11}^u]C_1'
-[C_{11}^u-2C_{44}^u+(r^u)^2C_{11}^u]C_2'
+2C_{44}^u s^u D_1'-2C_{44}^u s^u D_2'\nonumber\\
&+[C_{11}^d-2C_{44}^d-(r^d)^2C_{11}^d]C_4'
+2\iota s^d C_{44}^d D_4'=0\,,
\end{align}
\begin{align}
\label{eq:QRModBC6}
2C_{44}^u r^u C_1'-2C_{44}^u r^u C_2'+C_{44}^u[(s^u)^2-1]D_1'&+C_{44}^u[(s^u)^2-1]D_2'
\nonumber\\
&+2\iota r^d C_{44}^d C_4'+C_{44}^d[(s^d)^2+1]D_4'=0\,.
\end{align}
%%%%%%%%%%%%%%%%%%%%%%%%%%%%%%%%%%
\subsection{Dispersion relation}
\label{sec:dispersion}
%%%%%%%%%%%%%%%%%%%%%%%%%%%%%%%%%%
The six boundary conditions stated in equations~(\ref{eq:QRModBC1}), (\ref{eq:QRModBC2}), (\ref{eq:QRModBC3}), (\ref{eq:QRModBC4}), (\ref{eq:QRModBC5}) and (\ref{eq:QRModBC6})
form a linear system of six equations for six unknowns, $C_1'$, $C_2'$, $D_1'$, $D_2'$, $C_4'$ and $D_4'$\,.
For a nontrivial solution, the determinant of the coefficient matrix,~$M_{r}$\,, must be zero.
Upon factoring $\iota$ from the third and fourth rows, we write $M_{r}$ as
\begin{equation}
\label{eq:M_r}
\footnotesize{
\begin{bmatrix}
2 r^u\,{\rm e}^{\iota a'}&\,-2 r^u\,{\rm e}^{-\iota a'}&\,\dfrac{v_{r}^2\rho^u-2 C^u_{44}}{C^u_{44}}\,{\rm e}^{\iota b'}&\,\dfrac{v_{r}^2\rho^u-2 C^u_{44}}{C^u_{44}}\,{\rm e}^{-\iota b'}&\,0&\,0 \\ 
 \left[2 C_{44}^u-\rho^u v_{r}^2\right]\,{\rm e}^{\iota a'}& \, \left[2 C_{44}^u-\rho^u v_{r}^2\right]\,{\rm e}^{-\iota a'}&\,2C_{44}^us^u \,{\rm e}^{\iota b'}
&\,-2C_{44}^us^u \,{\rm e}^{-\iota b'}&\,0&\,0\\
1&\,1&\,s^u&\,-s^u&\,-1&\,\iota s^d\\
-r^u& \, r^u&\,1&\,1&\,-\iota r^d&\,-1\\
2 C_{44}^u-\rho^u v_{r}^2 &\,2 C_{44}^u-\rho^u v_{r}^2&\,2C_{44}^u s^u&\,-2C_{44}^u s^u&\,v_{r}^2\rho^d-2 C^d_{44}&\,2\iota C_{44}^d s^d\\
2C_{44}^{\mathstrut u} r^u&\,-2C_{44}^u r^u&\,v_{r}^2\rho^u-2 C^u_{44}&\,v_{r}^2\rho^u-2 C^u_{44}&\,2\iota C_{44}^d r^d&\, 2 C^d_{44}-v_{r}^2\rho^d
\end{bmatrix}},
\end{equation}
where we use
\begin{equation*}
C_{11}^u((r^u)^2+1)=C_{11}^u\frac{v_{r}^2}{(\alpha^u)^2}=\rho^u v_{r}^2\,\,,\quad (s^u)^2-1=\frac{v_{r}^2\rho^u-2 C^u_{44}}{C^u_{44}}\,,
\end{equation*}
\begin{equation*}
C^d_{11}(1-(r^d)^2)-2 C^d_{44}=v_{r}^2\rho^d-2 C^d_{44}\,\,,\quad C^d_{44}((s^d)^2+1)=2 C^d_{44}-v_{r}^2\rho^d\,.
\end{equation*}
Following a laborious process, shown in~\ref{app:disprel}, we obtain the determinant of the coefficient matrix,
\begin{equation}
\label{eq:ourdet}
\mathrm{D} := \det[M_r] =
4C^u_{44} \det\left[
\begin{array}{cc} 
s^u X & s^u S\\r^u T & r^u Y
\end{array}
\right]\,,
\end{equation}
where $X,Y,S,T$ are
\begin{equation*}
X :=
\left[(s^u)^2)-1\right]\left[-(v_{r}^2 q+ 2 p) B' +2 p r^d \cos b'\right] +
2 \left[r^u ( 2 p -v_{r}^2 \rho^d)\sin a' + r^d( 2 p + v_{r}^2 \rho^u)\cos a'\right],
\end{equation*}
\begin{equation*}
Y := \left[(s^u)^2)-1\right]\left[(v_{r}^2 q+ 2 p) A' -2 p s^d \cos a'\right] +
2 \left[- s^d ( 2 p +v_{r}^2 \rho^u)\cos b' - s^u( 2 p - v_{r}^2 \rho^d)\sin b'\right],
\end{equation*}
\begin{equation*}
S := \left[(s^u)^2)-1\right]\left[-(v_{r}^2\rho^u+ 2 p)s^d B' + (2 p-v_{r}^2 \rho^d) \cos b'\right] +
2 \left[ ( 2 p +v_{r}^2 q)\cos a' + 2 p r^u s^d \sin a'\right],
\end{equation*}
\begin{equation*}
T := \left[(s^u)^2)-1\right]\left[r^d (v_{r}^2 \rho^u + 2 p) A' -(2 p-v_{r}^2 \rho^d) \cos a'\right] - 
2 \left[( 2 p +v_{r}^2 q)\cos b' +2  s^u r^d p \sin b'\right],
\end{equation*}
\begin{equation*}
q:=\rho^u-\rho^d\,,\quad 
p:=C^d_{44}-C^u_{44}\,,\quad 
a':=\kappa r^u Z\,,\quad 
b':=\kappa s^u Z\,,
\end{equation*}
\begin{equation*}
A':=
\begin{dcases}
\frac{\sin a'}{r^u}&r^u\neq0\\
\kappa Z&r^u=0
\end{dcases}\,,
\qquad
B':=
\begin{dcases}
\frac{\sin b'}{s^u}&s^u\neq0\\
\kappa Z&s^u=0
\end{dcases}\,,
\end{equation*}
where both $A'$ and $B'$ are real, regardless of whether or not $r^u$ and $s^u$ are real or imaginary.
Both $A'$ and $B'$ are equal to $\kappa Z$\,, for $r^u=0$ and $s^u=0$\,, in the limit sense, in accordance with de l'H\^opital's rule.
Therefore, $X,Y,S,T$ are real, $\mathrm{D}=0$\,, for $s^u=0$ or $r^u=0$\,, and  whether $\mathrm{D}$ is real or imaginary depends only on the product~$s^u\,r^u$\,.
Thus, for $\alpha^d > \beta^d > v_{r}> \alpha^u > \beta^u$\,, the determinant is real, for $\alpha^u>v_{r} > \beta^u$\,, it is imaginary and, for $ \beta^u>v_r$\,, it is real again.

There are body-wave solutions for $r^u=0$\,, which means that $v_r=\alpha^u$\,, and for $s^u=0$\,, which means that $v_r=\beta^u$\,.
However, from analyses of equations~(\ref{eq:QRModBC1})--(\ref{eq:QRModBC4}), (\ref{eq:QRModBC5}) and~(\ref{eq:QRModBC6}), we conclude that the displacements are zero, and hence, these solutions are trivial.

We set $\mathrm{D}=0$ and solve numerically for $v_{r}$ for particular values of $C_{11}^u$\,, $C_{44}^u$\,, $\rho^u$\,, $C_{11}^d$\,, $C_{44}^d$\,, $\rho^d$\,, $Z$ and $\omega$\,.
Note that $a$\,, $a'$\,, $b$\,, $b'$ depend on $v_{r}$ and on $\kappa Z=\omega Z/v_r$\,, and $r^u$\,, $r^d$\,, $s^u$\,, $s^d$ depend on $v_{r}$\,.
Since the matrix includes terms depending on $\omega$\,, the quasi-Rayleigh waves---like Love waves and unlike standard Rayleigh waves---are dispersive.  

%%%
\begin{figure}
\begin{center}
\includegraphics[scale=0.5]{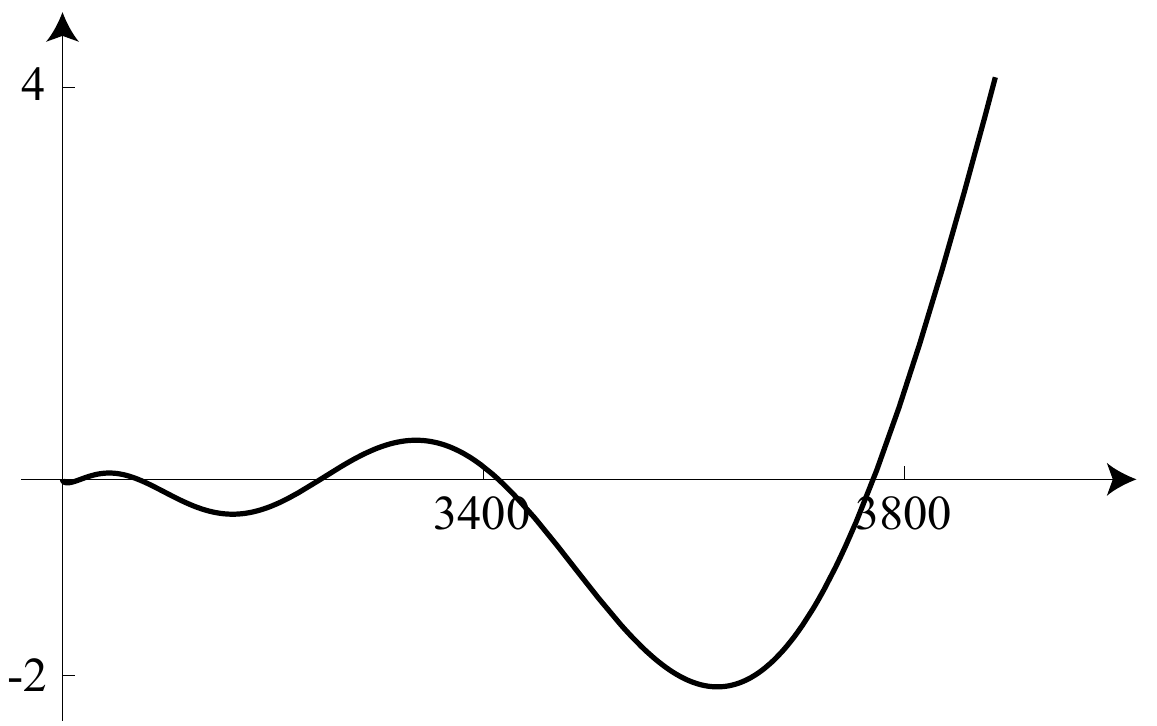}\hspace*{1in}\includegraphics[scale=0.5]{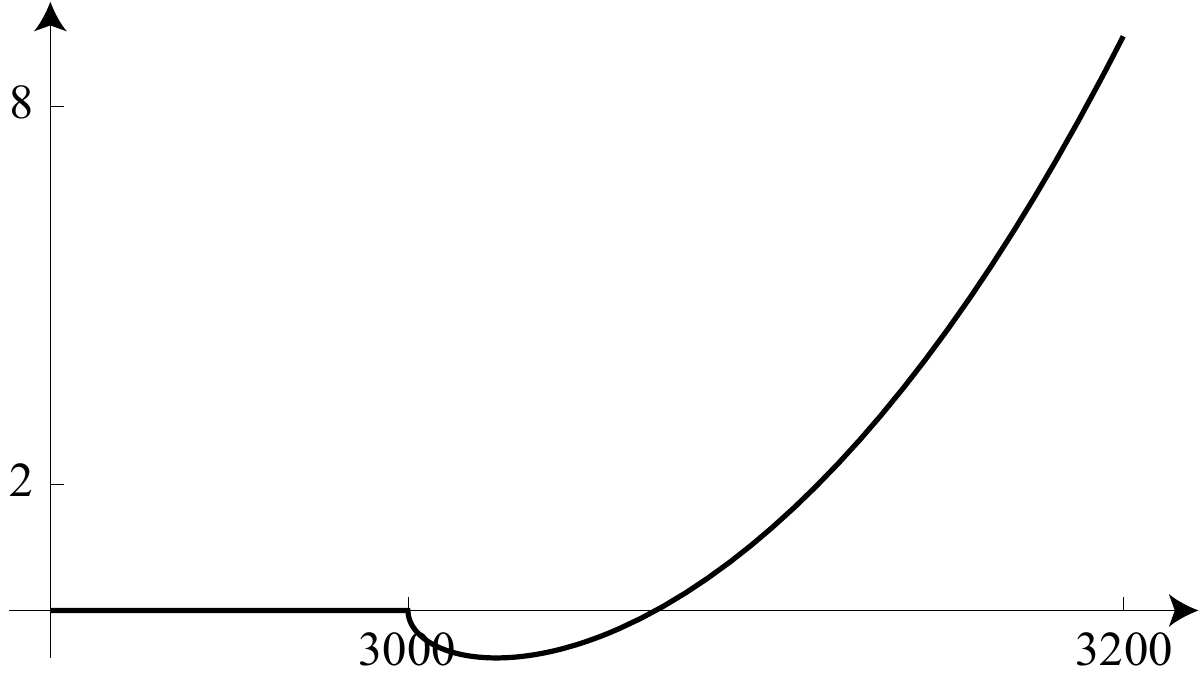}
\end{center}
\caption{\small{$\rm{Re}(\det[M_{r}])$ as a function of $v_{r}$\,. On the left, for $\omega=100~{\rm s}^{-1}$\,, there are five roots: $v_{r}=3017~{\rm m}/{\rm s}$\,,  
$v_{r}=3072~{\rm m}/{\rm s}$\,, $v_{r}=3245~{\rm m}/{\rm s}$\,,
$v_{r}=3414~{\rm m}/{\rm s}$ and $v_{r}=3771~{\rm m}/{\rm s}$\,.
On the right, for $\omega=5~{\rm s}^{-1}$\,, there is one root:
$v_{r}=3069~{\rm m}/{\rm s}$\,.
The values on the vertical axes are to be multiplied by $10^{32}$ and $10^{30}$\,, on the left- and right-hand sides, respectively.}}
\label{fig:real}
% from (left) 20160127qray copy2.nb and (right) 20160128qray.nb
\end{figure}

%%%
\begin{figure}
\begin{center}
\includegraphics[scale=0.5]{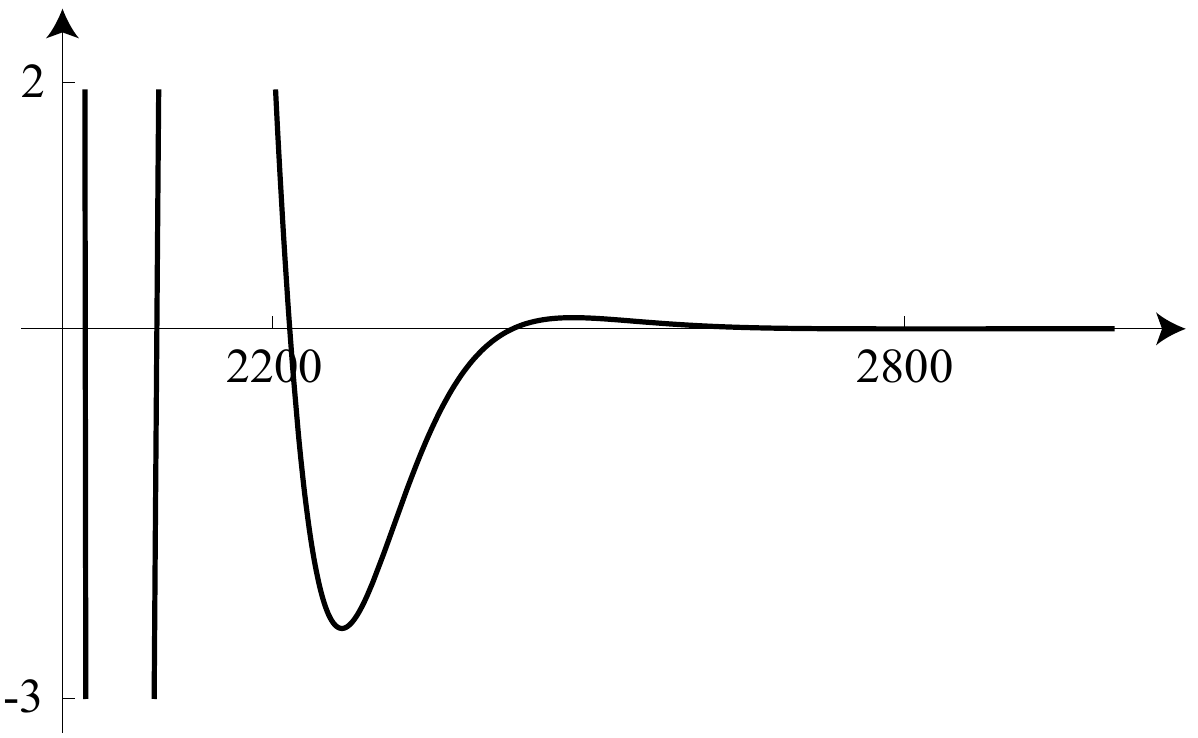}\hspace*{1in}\includegraphics[scale=0.5]{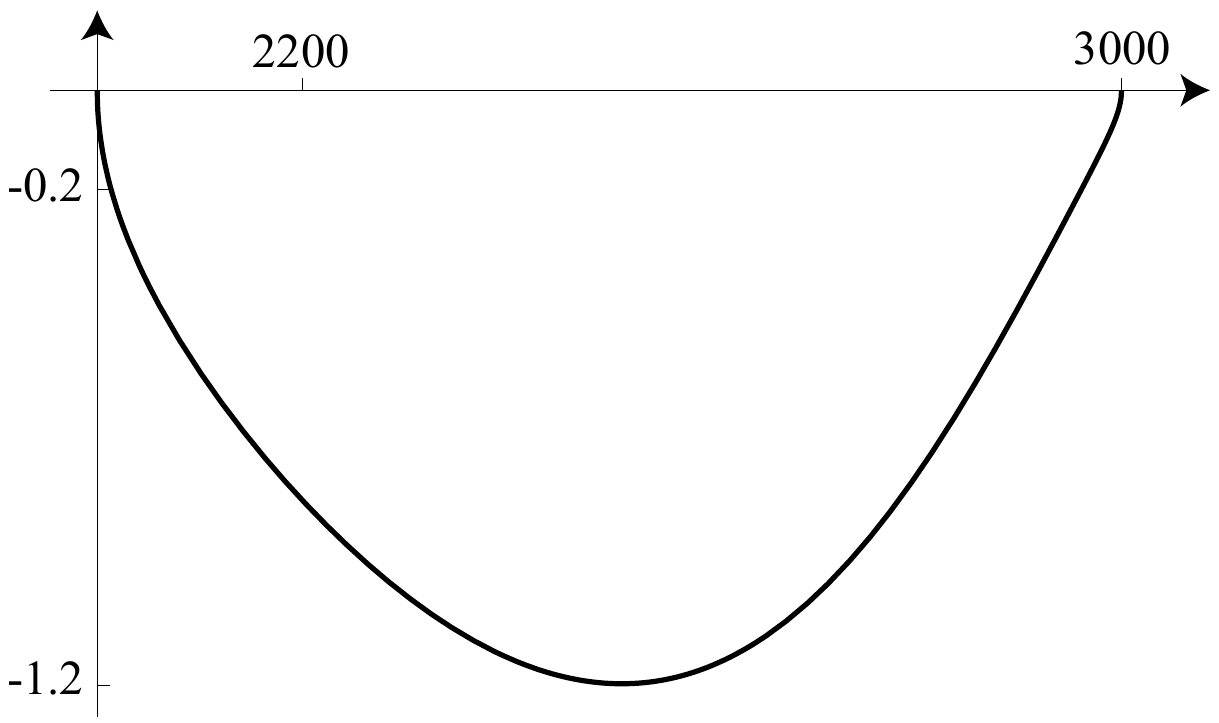}
\end{center}
\caption{\small{$\rm{Im}(\det[M_{r}])$ as a function of $v_{r}$\,.
On the left, for $\omega=100~{\rm s}^{-1}$\,, there are five roots:
$v_{r}=2022~{\rm m}/{\rm s}$\,,
$v_{r}=2090~{\rm m}/{\rm s}$\,, $v_{r}=2216~{\rm m}/{\rm s}$\,,
$v_{r}=2428~{\rm m}/{\rm s}$ and $v_{r}=2762~{\rm m}/{\rm s}$
\,.
On the right, for $\omega=5~{\rm s}^{-1}$\,, there are no roots, other than the trivial ones at the endpoints.
The values on the vertical axes are to be multiplied by $10^{37}$ and $10^{31}$\,, on the left- and right-hand sides, respectively.}}
\label{fig:imag2}
% from (left) 20160127qray copy2.nb and (right) 20160128qray.nb
\end{figure}

%%%
\begin{figure}
\begin{center}
\includegraphics[scale=0.5]{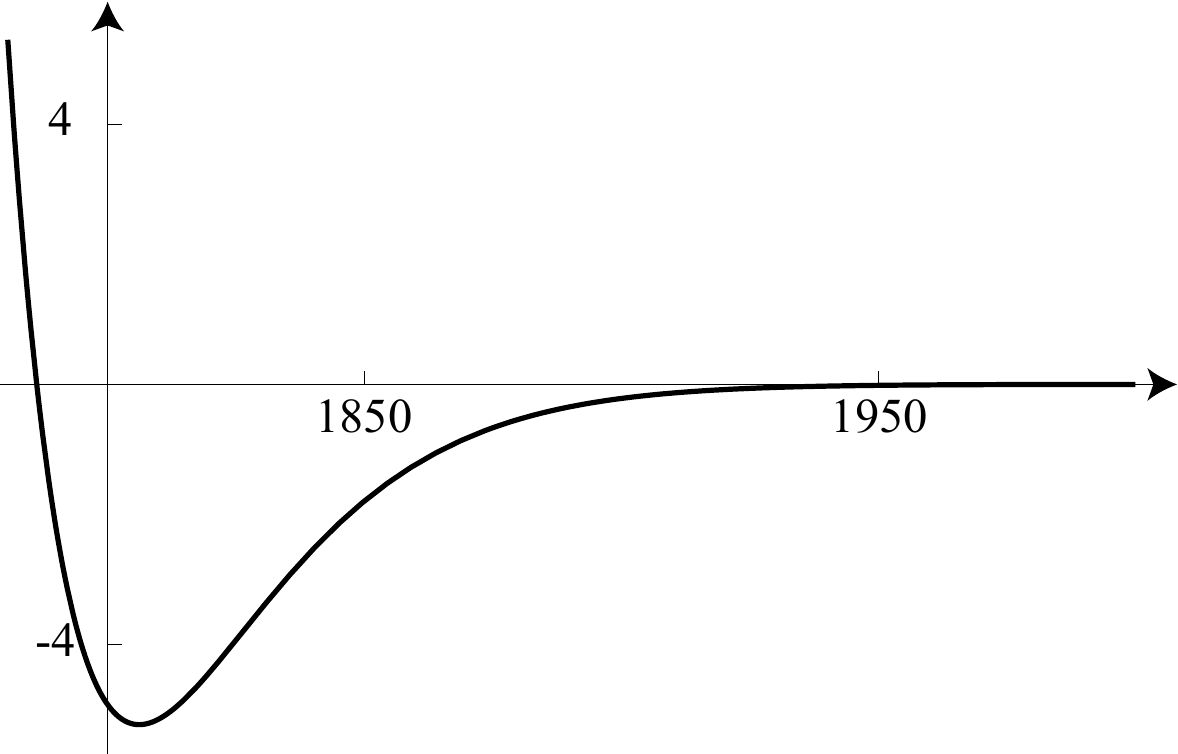}\hspace*{1in}\includegraphics[scale=0.5]{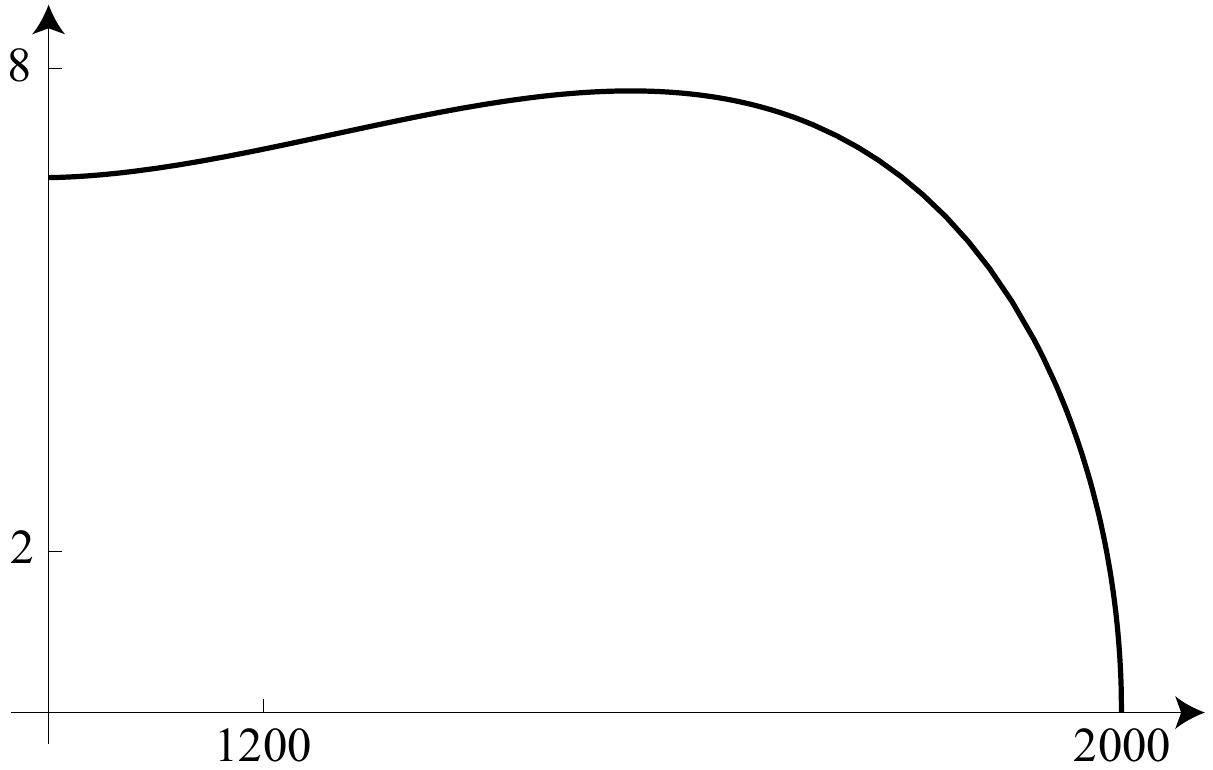}
\end{center}
\caption{\small{$\rm{Re}(\det[M_{r}])$ as a function of $v_{r}$\,.
On the left, for $\omega=100~{\rm s}^{-1}$\,, there is one root: $v_{r}=1786~{\rm m}/{\rm s}$\,, and a trivial root at the right endpoint.
On the right, for $\omega=5~{\rm s}^{-1}$\,,
there are no roots other than the trivial one at the right endpoint.
The values on the vertical axes are to be multiplied by $10^{43}$ and $10^{30}$\,, on the left- and right-hand sides, respectively.}}
\label{fig:real4}
% from (left) 20160127qray copy2.nb and (right) 20160128qray.nb
\end{figure}
%%%

The existence of Love waves requires $\beta^d > v_\ell  > \beta^u$, and of quasi-Rayleigh waves, $\alpha^d > \beta^d > v_{r} $, but~\citet{Udias1999} states that for the fundamental mode for high frequency, $v_r<\beta^u$,  but for higher modes, $v_r>\beta^u$\,.  So  it is not necessary that $v_r>\beta^u$, except for higher modes; a mode is a solution curve $v_r(\omega)$\,, and
the fundamental mode has content at all frequencies whereas higher modes
have cutoff frequencies below which they have no content.
Also, usually---but not always---\,$v_{r} > \alpha^u$~\citep{Udias1999}.
If $v_{r} <  \alpha^u$, then, from equations~(\ref{eq:QRGenSolAu}), (\ref{eq:QRuU1}) and~(\ref{eq:QRuU3}), there is a partially exponential variation in the layer instead of a purely sinusoidal variation.
If $\alpha^u \geq \beta^d$, $\det[M_{r}]$ is purely imaginary and there is still a solution for $v_{r}$ in $\det[M_r]=0$\,.

In Figures~\ref{fig:real} to~\ref{fig:real4}, we use the same parameters as in Figure~\ref{fig:love}, with addition of parameters $C_{11}^u=1.98\times 10^{10}~{\rm N}/{\rm m}^2$ ($\alpha^u=3000~{\rm m}/{\rm s}$) and also $C_{11}^d=10.985\times 10^{10}~{\rm N}/{\rm m}^2$ ($\alpha^d=6500~{\rm m}/{\rm s}$).
On the left-hand sides of Figures~\ref{fig:real} to~\ref{fig:real4}, we have
$\omega=100~{\rm s}^{-1}$ and, on the right-hand sides, we have $\omega=5~{\rm s}^{-1}$\,.
As $\omega$ or $Z$ increases, the number of solutions for $v_{r}$ increases.
For $\alpha^u=3000~{\rm m}/{\rm s}<v_r<\beta^d$\,, the determinant is real but for $\beta^u<v_{r}<\alpha^u$\,, the determinant becomes purely imaginary, and for $v_r<\beta^u$\,, the determinant becomes real again.
%%%%%%%%%%%%%%%%%%%%%%%%%%%%
\subsection{Sensitivity of dispersion relation}
\begin{figure}
\begin{center}
\includegraphics[scale=0.5]{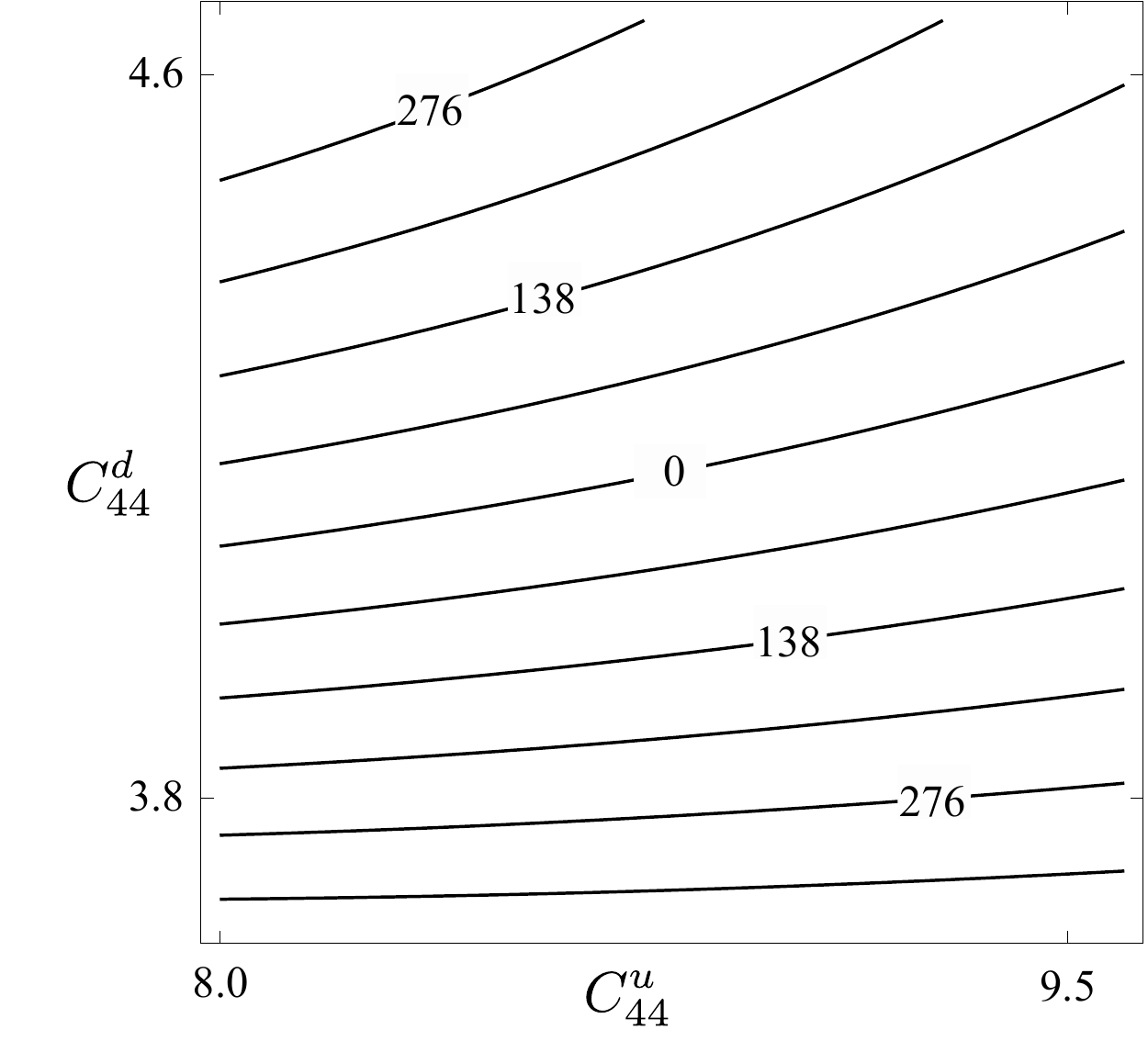}
\end{center}
\caption{\small{$\det[M_{r}]/10^{28}$ as a function of the elasticity parameters, $C_{44}^u$ and $C_{44}^d$\,, for $\omega=5~{\rm s}^{-1}$ and $v_{r}=3069~{\rm km}/{\rm s}$\,.
 The values on the horizontal and vertical axes are to be multiplied by $10^9$ and $10^{10}$\,, respectively.}}
 % from 20160413qray.nb
\label{fig:qraycontour1}
\end{figure}

%%%%%%%%%%%%%%%%%%%%%%%%%%%%
The sensitivity contour plot depicted in Figure~\ref{fig:qraycontour1} indicates that for the low frequency,~$\omega=5~{\rm s}^{-1}$\,, the determinant is sensitive to both $C_{44}^u$ and $C_{44}^d$\,.

\begin{figure}
\begin{center}
\includegraphics[scale=0.5]{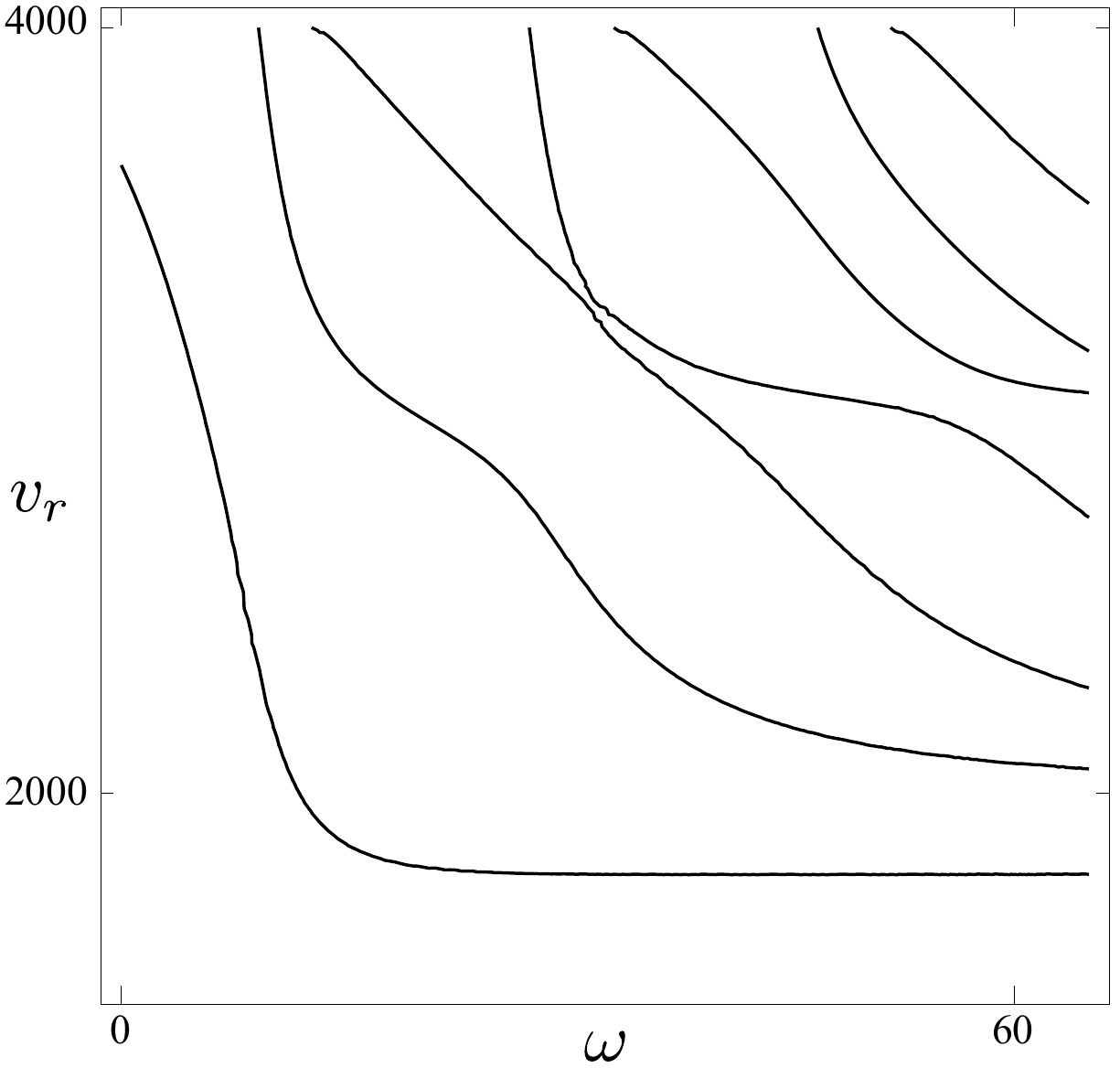}
\end{center}
\caption{\small{Zero lines of the quasi-Rayleigh-wave dispersion curves---$XY-ST=0$\,, in expression~(\ref{eq:ourdet})---as a function of speed,~$v_r$\,, and frequency,~$\omega$\,.}}
% from 20160413qray-rusu.nb
\label{fig:qraycontour5}
\end{figure}

In Figure~\ref{fig:qraycontour5}, we plot the dispersion curves for $XY-ST=0$\,.
This terms appears in expression~(\ref{eq:ourdet}) and is always real.
We do not plot $\mathrm{D}=0$ to avoid the trivial solutions, $r^u=0$ and $s^u=0$\,.
Also, in that manner, we avoid, for all frequencies, the transition from $\mathrm{D}$ being real for $v_r<\beta^u$,  being purely imaginary for $\beta^u<v_{r}<\alpha^u$ and being real for $\alpha^u<v_{r}<\beta^d$\,.
 
For the case of $v_r<\beta^u$\,, the fundamental mode still has a solution for $v_r$ for higher frequencies but the higher modes do not.
For high frequency, that fundamental-mode speed asymptotically approaches the classical Rayleigh wave speed in the layer, which is $0.89\beta^u$\,, and in the limit---as $\omega\rightarrow 0$---that fundamental mode, which---unlike the higher modes---has no low cutoff frequency speed, approaches the classical Rayleigh-wave speed in the halfspace, which is $0.91\beta^d$\,.%from rayleightest.nb
 
The dispersion curves in Figure~\ref{fig:qraycontour5} show one solution for $v_r=\beta^u=2000~{\rm m}/{\rm s}$ and an infinite number of solutions for $v_r=\alpha^u=3000~{\rm m}/{\rm s}$\,.
However, in both those cases, an analysis of equations~(\ref{eq:QRModBC1})--(\ref{eq:QRModBC4}), (\ref{eq:QRModBC5}) and~(\ref{eq:QRModBC6}) shows that the corresponding displacements are zero.
Hence, we refer to $r^u=0$\,, which means that $v_r=\alpha^u$\,, and to $s^u=0$\,, which means that $v_r=\beta^u$\,, as trivial solutions.
%%%%%%%%%%%%%%%%%%%%%%%%%%%%
\section{Comparison to literature results}
\label{sec:comparison}
%%%%%%%%%%%%%%%%%%%%%%%%%%%%
\citet{Love1911} considered the same problem but with the assumption of incompressibility.
\citet{Lee1932}, \mbox{\citet{Fu1946}} and~\citet{Udias1999} considered the same problem without the assumption of incompressibility but made simplifying assumptions before doing any calculations.

In \ref{app:comparison}, we translate the notations of~\citet{Love1911}, \citet{Lee1932} and~\citet{Fu1946} to our notation and find that the solutions match, within multiplicative factors.  
However, in the case of~\citet{Udias1999}, the following corrections are needed in his formu{l\ae}.
\begin{enumerate}
\item
The second term of formula~(10.87), $Zr'\sin a'$, should be $(Z/r')\sin a'$.
\item
In the third term of formula~(10.87), $\sin a'$ should be $\sin b'$.
\item
In the fourth term of formula~(10.86), $\sin b'$ should be $\cos b'$.
\item
The $(\beta')^2$ in the first denominator of formula~(10.92) should be $\beta^2$.
\item
Instead of $r$ and $s$,~\citet{Udias1999} should have $\overline{r}$ and $\overline{s}$, which are the magnitudes of $r$ and $s$\,.
\item
Our determinant, with the above corrections to~\citet{Udias1999}, is $4(C_{44}^u)^3 r^u s^u$ times his formula~(10.85); thus, formula~(10.85)  does not include solutions $r^u=0$ and $s^u=0$\,.
However, as we discuss in Section~\ref{sec:dispersion}, those solutions exhibit zero displacements, which might be the reason why \citet{Lee1932}, \citet{Fu1946} and \citet{Udias1999} omit them.
\end{enumerate}
%%%%%%%%%%%%%%%%%%%%%%%%%%%%
\section{Conclusions}
%%%%%%%%%%%%%%%%%%%%%%%%%%%%
In this paper, we obtain the dispersion relations for both Love and quasi-Rayleigh waves within the same model.
Both are guided waves in an elastic layer constrained by a vacuum and an elastic halfspace.
According to the calculations, and as illustrated in
Figures~\ref{fig:love3x3contourw5-2} and  \ref{fig:qraycontour5},
the speeds of these waves differ.
Thus---if measurable on a seismic record---their arrivals should be distinct, apart from the fact that their polarizations are orthogonal to each other.

The speeds of both Love and quasi-Rayleigh waves are obtained from their respective dispersion relations.
In the high-frequency example, the fundamental Love-wave mode has a speed that is slightly greater than the $S$-wave speed in the layer, and in the low-frequency example, slightly lower than the $S$-wave speed in the halfspace.

The first quasi-Rayleigh-wave mode exemplified on the left plot of Figure~\ref{fig:real}, in which the determinant is real and $v_{r} > \alpha^u$, has a speed that is slightly greater than the $P$-wave speed in the layer.
The first quasi-Rayleigh-wave mode exemplified on the left plot of Figure~\ref{fig:imag2}, in which the determinant is purely imaginary and $\beta^u < v_{r} < \alpha^u$, has a speed that is slightly greater than the $S$-wave speed in the layer. 
The highest-mode speeds of both the Love wave, in Figure~\ref{fig:love}, and the quasi-Rayleigh wave, in Figure~\ref{fig:real}, are less than the shear-wave speed in the halfspace.
For the low-frequency example in the right plot of Figure~\ref{fig:real}, there is only one quasi-Rayleigh wave mode.

The dispersion curves are given as the zero contours for Love waves, in Figure~\ref{fig:love3x3contourw5-2}, and for quasi-Rayleigh waves, in Figure~\ref{fig:qraycontour5}.
In the latter figure---and in agreement with Figure~10.14 of~\citet{Udias1999}---the fundamental mode has all frequencies, which means that it has no cutoff frequency.
For high frequency, the fundamental-mode speed asymptotically approaches the classical Rayleigh-wave speed in the layer, and---in the limit as $\omega\rightarrow 0$\,---that speed approaches the classical Rayleigh-wave speed in the halfspace.  
 
Upon deriving the dispersion relations for both the Love and the quasi-Rayleigh waves, we focus on the latter to give details of the expansion of the $6\times 6$ matrix and its determinant.
We compare our results to several past studies, including the one of~\citet{Love1911}, which assumes incompressibility, and the one of~\citet{Udias1999}, in which there appear errors in the equations, though not in the dispersion-curve plots.
Unlike \citet{Love1911}, who assumes incompressibility, \citet{Udias1999}, who assumes a Poisson's ratio of $1/4$\,, and~\citet{Fu1946}, who studies limiting cases, we do not make any simplifying assumptions prior to calculations.

In the context of the $6\times 6$ matrix, our plots demonstrate the sensitivity of the dispersion relations to elasticity parameters. 
Also, we show that the solutions $r^u=0$\,, which means that $v_r=\alpha^u$\,, and $s^u=0$\,, which means that $v_r=\beta^u$\,, have zero displacements and can be considered trivial solutions.
%%%%%%%%%%%%%%%%%%%%%%%%%%%%%%%
\section{Future work}
%%%%%%%%%%%%%%%%%%%%%%%%%%%%%%%
Given dispersion relations of the Love and the quasi-Rayleigh waves, we expect to invert the measurements of speed for elasticity parameters and mass densities of both the layer and the halfspace, as well as for the layer thickness.
Explicit expressions presented in this paper allow us to formulate the inverse problem and examine the sensitivity of its solution.

Also, we could formulate dispersion relations for the case of an anisotropic layer and an anisotropic halfspace.
Such a formulation would require modified boundary conditions and equations of motion.
Further insights into such issues are given by \citet{Babich}.

\citet[pp.~41--43, 62]{GrantWest1965} discuss the equations of motion in transversely isotropic media and reference~\citet{Sato1950}, who discusses Rayleigh waves on the surface of a transversely isotropic halfspace.
In a seismological context, a transversely isotropic layer could result from the~\citet{Backus1962} average of a stack of thin parallel isotropic layers.  
Notably,~\citet{Rudzki1912} already examined Rayleigh waves on the surface of a transversely isotropic halfspace.
%%%%%%%%%%%%%%%%%%%%%%%%%%%%%%%
\section*{Acknowledgments}
%%%%%%%%%%%%%%%%%%%%%%%%%%%%
We wish to acknowledge discussions with Tomasz Danek and Michael Rochester.
We also acknowledge the graphical support of Elena Patarini.
This research was performed in the context of The Geomechanics Project supported by Husky Energy. 
Also, this research was partially supported by the Natural Sciences and Engineering Research Council of Canada, grant 238416-2013.
\nocite{*}
\bibliographystyle{chicago}
\bibliography{qrayandlovebib}
%%%%%%%%%%%%%%%%%%%%%%%%%%%%%%%%
\setcounter{section}{0}
\setlength{\parskip}{0pt}
\renewcommand{\thesection}{Appendix~\Alph{section}}
%%%%%%%%%%%%%%%%%%%%%%%%%%%%%%%%
\section{Details of dispersion relation derivation}
\label{app:disprel}
%%%%%%%%%%%%%%%%%%%%%%%%%%%%%%%%
In this appendix, we present operations to compute the determinant of the $6\times 6$ matrix stated in expression~(\ref{eq:M_r}).
We invoke several algebraic properties that allow us to obtain a $2\times 2$ matrix.
In this process, we use the following notational abbreviations.
\begin{itemize}
	\item $C1,\ldots,C6$ denotes columns $1, \ldots,6$
	\item $R1,\ldots,R6$ denotes rows $1, \ldots,6$
	\item $C1\mapsto C1+C2$ denotes replacement of $C1$ by $C1+C2$\,, etc. 
	\item  $A':=\sin a'/r^u$
	\item $B':=\sin b'/s^u$
	\item $u_\beta:=v_{r}/\beta^u$
\end{itemize}
Using this notation, we perform the the following sequence of operations.\\
\begin{tabular}{p{8cm}p{8cm}}
\begin{enumerate}
\addtolength{\itemsep}{-2pt}
	\item Factor out $1/C^u_{44}$ from $R1$
	\item $C1\mapsto C1-C2$
	\item Factor out 2 from $C1$
	\item $C2\mapsto C2+C1$
	\item $C3\mapsto C3-C4$
	\item Factor out 2 from $C3$
	\item $C4\mapsto C3+C4$
	\item Factor out $r^u$ from $C1$ and $s^u$ from $C3$
        \item $C2\mapsto C2- C3$
\end{enumerate}
&
\begin{enumerate}
	\setcounter{enumi}{9}
	\item $C4\mapsto C4+C1$
	\item $R6\mapsto R6-2 C^u_{44} R4$
	\item $R5\mapsto R5-2 C^u_{44} R3$
	\item Factor out $\iota$ from $C2, C3, C5$
	\item Factor out $-\iota$ from $R3,R2, R5$
	\item Factor out $C^u_{44}$ from $ R1$  and $R2$
	\item Move $C3$ to the first column and shift the former first column and second column to the right; in other words, let $C1'=C1$, $C2'=C2$, $C1\mapsto C3$,  $C2\mapsto C1'$,  $C3\mapsto C2'$.
	\end{enumerate}
\end{tabular}
\begin{align*}
&\det[M_r]= 4r^u s^u  C^u_{44}\\
&\hspace{-1cm}\det\left[
{\scriptsize
\begin{array}{cccccc}
((u_\beta)^2-2) B' & 2 \cos a' & 2 (r^u)^2 A' -((u_\beta)^2-2)B' & ((u_\beta)^2-2)\cos b'+2 \cos a' & 0 &0\\ 
 2 \cos b'  &  ((u_\beta)^2-2) A'  & - 2 \cos b'-((u_\beta)^2-2)\cos a'  & ((u_\beta)^2-2)A' -2 (s^u)^2B' & 0 &0\\
1 & 0 & 0 & 0 & -1 &-s^d\\
0 &   -1 & 0 & 0  & - r^d &-1\\
0  & 0 &  - v_{r}^2\rho^u  & 0 &v_{r}^2\rho^d-2  (C^d_{44}-C^u_{44}) &  - 2 s^d(C^d_{44}-C^u_{44})\\
0  & 0 & 0 & v_{r}^2\rho^u & 2 r^d(C^d_{44}-C^u_{44}) &2 (C^d_{44}-C^u_{44})-v_{r}^2\rho^d   
\end{array}}
\right]\,.
\end{align*}
Let us write the above matrix in the block form, 
\begin{equation*}
\left[\begin{array}{cc} B & 0 \\ A & C\end{array}\right]\,,
\end{equation*}
where
\begin{equation*}
B:=\left[
\begin{array}{cccc}
((u_\beta)^2-2) B' &\quad 2 \cos a' &\quad 2 (r^u)^2 A' -((u_\beta)^2-2)B' &\quad  ((u_\beta)^2-2)\cos b'+2 \cos a'\\ 
 2 \cos b'  &\quad  ((u_\beta)^2-2) A'  &\quad - 2 \cos b'-((u_\beta)^2-2)\cos a'  &\quad ((u_\beta)^2-2)A' -2 (s^u)^2B'
\end{array}\right]\,,
\end{equation*}
\begin{equation*}
A:=\left[\begin{array}{cccc}1 & 0 & 0 & 0  \\
0 &   -1 & 0 & 0 \\
0  & 0 &  - v_{r}^2\rho^u  & 0 \\
0  & 0 & 0 & v_{r}^2\rho^u   
\end{array}\right]\,,
\qquad
C:=\left[\begin{array}{cc}
-1 & - s^d\\
 - r^d & - 1\\
v_{r}^2\rho^d-2  (C^d_{44}-C^u_{44}) & - 2 s^d(C^d_{44}-C^u_{44}) \\
 2 r^d(C^d_{44}-C^u_{44})   & 2 (C^d_{44}-C^u_{44})-v_{r}^2\rho^d 
\end{array}\right]\,.
\end{equation*}
Since $A$ is invertible, we have
\begin{equation*}
\left[ \begin{array}{cc} B & 0 \\ A & C\end{array}\right]=\left[ \begin{array}{cc} 0 & I_2 \\ I_4 & 0\end{array}\right] \left[ \begin{array}{cc} A & C \\ B & 0\end{array}\right]=
\left[ \begin{array}{cc} 0 & I_2 \\ I_4 & 0\end{array}\right] \left[ \begin{array}{cc} A & 0 \\ 0 & I_2\end{array}\right] \left[ \begin{array}{cc} I_4 & 0 \\B  & -I_2\end{array}\right]
\left[ \begin{array}{cc} I_4 & A^{-1} C  \\ 0 & BA^{-1} C\end{array}\right]\,,
\end{equation*}
and, hence,
\begin{equation*}
\det\left[ \begin{array}{cc} B & 0 \\ A & C\end{array}\right]=\det[A]\det[B A^{-1}C]\det \left[ \begin{array}{cc} 0 & I_2 \\ I_4 & 0\end{array}\right] 
\det \left[ \begin{array}{cc} I_4 & 0 \\ B  & -I_2\end{array}\right]=\det[A]\det[B A^{-1}C]\,.
\end{equation*}
Finally, we can write the determinant of the coefficient matrix, $M_r$\,, as
\begin{equation*}
\det[M_r]=4r^u s^u  C^u_{44}(XY -ST)=4C^u_{44} \det\left[\begin{array}{cc} s^u X & s^u S\\r^u T & r^u Y\end{array}\right]\,,
\end{equation*}
where $X, Y, S, T$ are given by formu{l\ae} below equation (\ref{eq:ourdet}).
%%%%%%%%%%%%%%%%%%%%%%%%%%%%%%%%
\section{Notational differences within literature}
\label{app:comparison}
%%%%%%%%%%%%%%%%%%%%%%%%%%%%%%%%
In this appendix, we present the notational translations for the quasi-Rayleigh waves among works discussed in Section~\ref{sec:comparison}, and compare differences found therein.
Herein, for convenience, we use definitions~(\ref{eq:QRConvI}) and (\ref{eq:QRConvII}).
%%%%%%%%%%%%%%%%%%%%%%%%%%%%%%%%
\subsection{Comparison with~\citet{Love1911}}
%%%%%%%%%%%%%%%%%%%%%%%%%%%%%%%%
\citet{Love1911} simplifies his formulation by assuming incompressibility, which implies that $\alpha^u\rightarrow\infty$ and $\alpha^d\rightarrow\infty$\,.
As a consequence, $r^u=\iota$ and $r^d=1$\,. 
It follows that
\begin{equation*}
\cos(a')=\cos(\kappa r^uZ)=\cos(\iota\kappa Z)=\cosh(\kappa Z)\,,
\end{equation*}
\begin{equation*}
\sin(a')=\sin(\kappa r^uZ)=\sin(\iota\kappa Z)=\iota\sinh(\kappa Z)\,,
\end{equation*}
\begin{equation*}
\cosh(\iota\kappa s^uZ)=\cosh(\iota b')=\cos(b')\,,
\end{equation*}
\begin{equation*}
\sinh(\iota\kappa s^uZ)=\sinh(\iota b')=\iota\sin(b')\,.
\end{equation*}
Consequently, his
\begin{equation*}
\xi\eta'-\xi'\eta=0
\end{equation*}
becomes
\begin{equation*}
\frac{\iota s^u}{(C_{44}^u)^2}(XY-ST)=\frac{\iota\,\mathrm{D}}{4(C_{44}^u)^3\,r^u}=0\,,
\end{equation*}
in our notation, where $\rm D$ is defined by expression~(\ref{eq:ourdet}).
Under the assumption of incompressibility, $r^u$ becomes $\iota$\,, and thus
\begin{equation*}
\xi\eta'-\xi'\eta=\frac{\mathrm{D}}{4(C_{44}^u)^3}\,.
\end{equation*}
%%%%%%%%%%%%%%%%%%%%%%%%%%%%%%%%
\subsection{Comparison with~\citet{Lee1932}}
%%%%%%%%%%%%%%%%%%%%%%%%%%%%%%%%
\citet{Lee1932} obtains a determinantal equation in terms of $\sin$ and $\cos$\,.
In our notation, his determinantal equation,
\begin{equation*}
\xi\eta'-\xi'\eta=0\,,
\end{equation*}
becomes
\begin{equation*}
\frac{XY-ST}{(C_{44}^u)^2}=\frac{\mathrm{D}}{4r^us^u(C_{44}^u)^3}=0\,.
\end{equation*}
In other words, our expressions differ by a multiplicative factor.
%%%%%%%%%%%%%%%%%%%%%%%%%%%%%%%%
\subsection{Comparison with~\citet{Fu1946}}
%%%%%%%%%%%%%%%%%%%%%%%%%%%%%%%%
\citet{Fu1946} obtains a determinantal equation in terms of $\sinh$ and $\cosh$\,.
Invoking standard expressions,
\begin{equation*}
\cosh(\iota a')=\cos(a')\,,\quad\cosh(\iota b')=\cos(b')\,,\quad
\sinh(\iota a')=\iota\sin(a')\,,\quad\sinh(\iota b')=\iota\sin(b')\,,
\end{equation*}
we write his determinantal equation,
\begin{equation*}
\zeta\eta'-\zeta'\eta=0\,,
\end{equation*}
as
\begin{equation*}
(XY-ST)\,\kappa^8=\frac{\kappa^8\,\mathrm{D}}{4\,C_{44}^u\,r^u\,s^u}=0\,,
\end{equation*}
in our notation.
%%%%%%%%%%%%%%%%%%%%%%%%%%%%%%%%
\subsection{Comparison with~\citet{Udias1999}}
%%%%%%%%%%%%%%%%%%%%%%%%%%%%%%%%
Referring to \citet{Udias1999}, his determinantal equation,
\begin{equation*}
\xi\eta'-\xi'\eta=0\,,
\end{equation*}
becomes
\begin{equation*}
\frac{XY-ST}{(C_{44}^u)^2}=\frac{\mathrm{D}}{4\,(C_{44}^u)^3\,r^u\,s^u}=0\,,
\end{equation*}
where we also use the six corrections listed in Section~\ref{sec:comparison}.
%%%%%%%%%%%%%%%%%%%%%%%%%%%%%%%%
\end{document}